\documentclass[article,prc,aps,epsfig,floatfix,twocolumn]{revtex4}
\usepackage{amssymb}
\usepackage{graphicx,psfrag}

\providecommand{\tabularnewline}{\\}

\begin{document}
\newcommand{\code}{$\mathtt{FREYA}$}
\newcommand{\beq}{\begin{eqnarray}}
\newcommand{\eeq}{\end{eqnarray}}
\newcommand{\SKIP}[1]{}
\newcommand{\nubar}{{\overline{\nu}}}

\title{Event-by-event study of prompt neutrons from $^{239}\textrm{Pu}(n,f)$}

\author{R.~Vogt$^{1,2}$, J.~Randrup$^3$, J.~Pruet$^1$, and W.~Younes$^1$}

\affiliation{
$^1$Physics Division, Lawrence Livermore National Laboratory, Livermore, CA
94551, USA\break
$^2$Physics Department, University of California, Davis, CA 95616, USA\break
$^3$Nuclear Science Division, Lawrence Berkeley National Laboratory, 
Berkeley, CA 94720, USA}

\date{\today}

\begin{abstract}
Employing a recently developed Monte-Carlo model,
we study the fission of $^{240}$Pu induced by neutrons with
energies from thermal to just below the threshold for second chance fission.
Current measurements of the mean number of prompt neutrons emitted in fission, 
together with less accurate measurements of the neutron energy spectra, 
place remarkably fine constraints on predictions of microscopic calculations. 
In particular, the total excitation energy of the nascent fragments must be 
specified to within 1 MeV to avoid disagreement with measurements of
the mean neutron multiplicity.
The combination of the Monte-Carlo fission model 
with a statistical likelihood analysis
also presents a powerful tool for the evaluation of fission neutron data.
Of particular importance is the the fission spectrum,
which plays a key role in determining reactor criticality.
We show that our approach can be used to develop an estimate of the
fission spectrum with uncertainties several times smaller
than current experimental uncertainties for outgoing neutron energies of
less than 2 MeV.
\end{abstract}

\maketitle

\section{Introduction}

The quest for a fundamental theory of fission began with the 1939
seminal work of Bohr and Wheeler \cite{BohrWheeler}, 
the same year as this phenomenon was discovered by Hahn and Strassmann 
\cite{HahnStrassmann} and interpreted by Meitner and Frisch \cite{Meitner39}.
Bohr and Wheeler used the liquid-drop model to make predictions
that were remarkably realistic given the paucity of available data.
The current theoretical descriptions of fission reflect the
complexity and richness revealed over 70 years of experimental studies,
emphasizing the multi-dimensional, dynamic, and microscopic aspects.
In particular, a refined version of the liquid drop model that
includes a finite interaction range and quantum shell corrections 
has formed the basis for extensive calculations of 
the potential-energy surfaces associated with the multidimensional shape
of fissioning nuclei 
(see Refs.\ \cite{MMSI-Nature409,MollerPRC79} and references therein). 
Concurrently, a program is underway to develop a fully microscopic treatment 
of fission in terms of a quantum many-body treatment of protons and neutrons
subject to an adjustable effective (in-medium) interaction
 \cite{BergerNPA428,GouttePRC71,DubrayPRC77}.

Despite the many theoretical advances, there is not yet a 
quantitative theory of fission.  This is unfortunate because nuclear
fission remains important to society at large due to its many practical
applications, including energy production and security.  For example, 
reactors and other critical systems demand that neutron growth be known to 
about the 0.1$\%$ level for model simulations to be reliable.  In such cases, 
scattering experiments are insufficiently accurate,  requiring reliance on
more inclusive, higher statistics integral critical assembly experiments. 

Furthermore, in the last few years efforts have been underway 
to develop systems capable of detecting concealed nuclear material.  
These applications 
place entirely different demands on fission models by attempting to exploit
specific information carried by particles resulting from fission.  
Thus there is a need for a fission description that accounts for 
particle correlations and fluctuations on an event-by-event level.  
Such a description, 
employing a model incorporating the relevant physics with a few key parameters,
compared to the pertinent data through a statistical analysis, 
presents a potentially powerful tool for bridging the gap between 
current microscopic models and important fission observables 
and for improving estimates of the relatively gross fission characteristics 
important for applications.  
This type of approach also provides a means of using 
readily measured observables to constrain our understanding 
of the microscopic details of fission.

Relatively recently, Lemaire {\it et al.}\ \cite{LemairePRC72} implemented
a Monte-Carlo simulation of fission fragment statistical decay
by sequential neutron emission for spontaneous fission of $^{252}{\rm Cf}$ 
and thermal fission of $^{235}{\rm U}$.
That work demonstrated how fission event simulations, 
in conjunction with experimental data on fission neutrons 
and physics models of fission and neutron emission, 
can be used to predict the neutron spectrum 
and to validate and improve the underlying physics models.

In the present work, we have implemented a conceptually similar approach
and applied it to calculate the sequential neutron emission 
for the neutron induced fission of $^{240}\textrm{Pu}$. 
Specifically, we have adapted the recently developed 
fission event generation model \code\ \cite{RV}
to calculate the production and decay of fission fragments
and used maximum-likelihood analysis to estimate properties of the
emitted fission neutrons and their correlation coefficients. 
To our knowledge, such correlations have not been extracted before for
fission neutrons in a physics-based Monte-Carlo simulation. The detailed
statistical analysis presented here is essential for developing a more
quantitative understanding of fission
and obtaining better evaluations of fission data for various applications.

First, in Sect.\ \ref{method},
we present the framework for the statistical analysis employed
for obtaining estimates of the model parameters and the neutron observables,
as well as the correlations between the various quantities of interest. 
We then discuss in Sect.\ \ref{data} the experimental data used in this work
with a particular emphasis on experimental uncertainties. 
Subsequently, in Sect.\ \ref{model},
we describe the physics ingredients of the \code\ simulations.
Finally, in Sect.\ \ref{results} we present calculated results for
the $^{239}\textrm{Pu}(n,f)$ neutron spectrum and other observables
for incident neutron energies, $E_n$, from 0.5 to 5.5 MeV.

\section{Statistical method}
\label{method}

Here we briefly describe the statistical method used
for determining model parameters and reaction observables.

There are a number of different techniques for estimating model parameter
values and although their relative merits are being vigorously debated
they often differ very little in their actual results.
Our present analysis is inspired by the general inverse problem theory
developed by Tarantola~\cite{tarantola}.

We introduce a number of model parameters $\{\alpha_k\}$ (defined in 
Sect.\ \ref{model}).  Since the theory does not, a priori, specify the
parameter values, we assume that the parameter values are uniformly distributed
over a reasonable interval in parameter space.  For a specified set of 
parameter values $\{\alpha_k^{(m)}\}$, we generate a large sample
of fission events from which we then extract the particular observables of
interest, $\{{\cal C}_i\}$.
These calculated values are then compared with the corresponding
experimental values, $\{{\cal E}_i\}$.

Specifically, for each parameter set $\{\alpha_k^{(m)}\}$ we calculate the
$\chi^2$ deviation of the calculated observables
from their measured values,
\beq
\chi^2_m\ \equiv\ \chi^2\{\alpha_k^{(m)}\}\ \equiv\ 
\sum_i {({\cal C}_i\{\alpha_k^{(m)}\}-{\cal E}_i)^2  \over \sigma_i^2}\ .
\eeq
Here $\{\sigma_i\}$ are the uncertainties 
in the experimental values 
Division by these quantities ensures that well-measured observables 
carry more weight than those that are poorly measured.

The key feature of the method \cite{tarantola} is that a likelihood
is assigned to each particular 
set $m$ of model parameter values based on how well 
the corresponding model calculation reproduces the experimental results,
\beq
w_m\ \equiv\ w\{\alpha_k^{(m)}\} \propto\ 
{\rm e}^{-\mbox{$1\over2$}\chi^2\{\alpha_k^{(m)}\}}\ .
\eeq
This quantity is then taken as the relative probability
that those parameter values are the ``correct'' ones.
In this manner, one may define a probability density 
in the space of model parameters, $P\{\alpha_k\}\equiv w\{\alpha_k\}/W$,
where $W\equiv\sum_m w_m$ is the sum of all the weights.

Once the probability density of model parameter values has been obtained,
their corresponding statistical distribution of the observables 
can readily be calculated. 
Thus the {\em best estimate} for the model parameter values, 
$\{\tilde{\alpha}_k\}$, is given by the likelihood-weighted average,
\beq
\tilde{\alpha}_k\ \equiv\ \prec \alpha_k \succ\ 
\equiv\ {1\over W}\sum_m w_m \alpha_k^{(m)}\ 
\approx\ \alpha^0_k\ .
\eeq
The last relation indicates that the best estimate is approximately equal to 
the {\em most likely} value $\alpha_k^0$, 
{\em i.e.}\ the value having the largest likelihood.
The covariances among the parameter values are similiarily calculated, 
\beq\label{sigmakk}
\tilde{\sigma}_{kk'}\ \equiv\ 
\prec(\alpha_k-\tilde{\alpha}_k)(\alpha_{k'}-\tilde{\alpha}_{k'})\succ\ .
\eeq
The diagonal elements, $\tilde{\sigma}_{kk}=\tilde{\sigma}_k^2$, are the usual
variances with $\tilde{\sigma}_k$ the standard deviations of the parameter 
values and represent the squares of the uncertainties 
on the values of the individual model parameter $\alpha_k$.
The off-diagonal elements give the covariances between
two model parameters.
It is often more instructive to employ the associated 
{\em correlation coefficients},
$C_{kk'}\equiv\tilde{\sigma}_{kk'}/[\tilde{\sigma}_k\tilde{\sigma}_{k'}]$.

\begin{figure}[b]	
\includegraphics[width=3.1in]{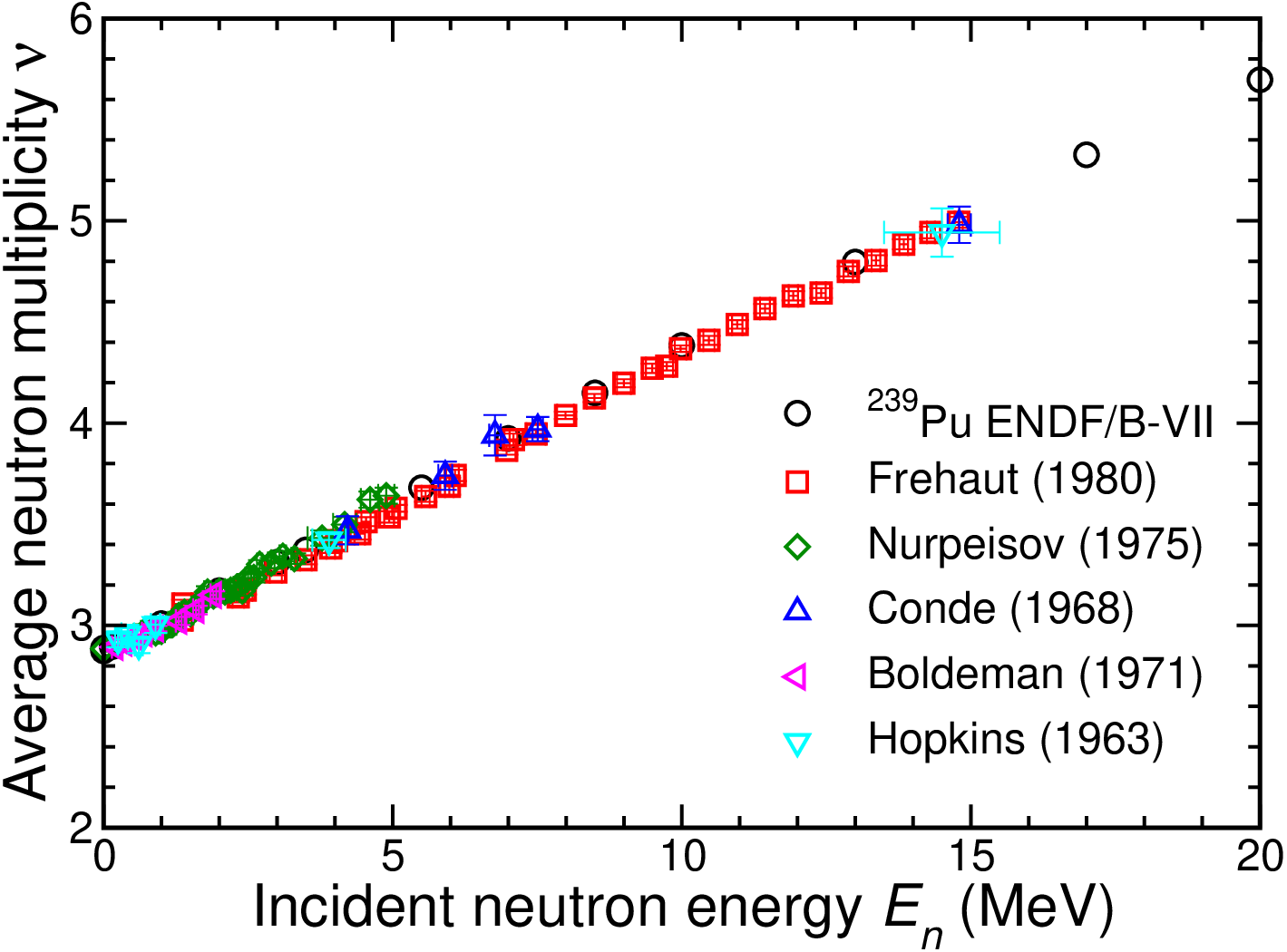}
\caption{(Color online)
The evaluated ENDF/B-VII data for the average prompt neutron multiplicity 
$\nubar$ as a function of the incoming neutron energy $E_{n}$,
together with the experimental data from 
Refs.~\protect\cite{FrehautNu,NurpeisovNu,CondeNu,BoldemanNu,HopkinsNu}.}
\label{f:nuBarFigure}
\end{figure}		

An analogous procedure can be carried out to obtain
best estimates for the various calculated quantities,
{\em i.e.}\ for the observables $\{{\cal C}_i\}$.
Thus, if ${\cal C}_i^{(m)}\equiv{\cal C}_i\{\alpha_k^{(m)}\}$
denotes the value of ${\cal C}_i$
calculated with the particular parameter values $\{\alpha_k^{(m)}\}$,
then the best estimate for the observable ${\cal C}_i$ is given by
\beq
\tilde{\cal C}_i\ \equiv\ \prec {\cal C}_i\succ\
=\ {1\over W}\sum_m w_m {\cal C}_i^{(m)}\ 
\approx\ {\cal C}_i\{\alpha^0_k\}\ .
\eeq
The last relation expresses the fact that the best estimate
is approximately equal to the most likely result,
{\em i.e.}\ the result obtained with the most likely parameter values.

Covariances between different observables, $\{{\cal C}_i\}$, are calculated as
\beq\label{sigmaij}
\tilde{\sigma}_{ij}\ \equiv\ 
\prec({\cal C}_i-\tilde{\cal C}_i)({\cal C}_j-\tilde{\cal C}_j)\succ
\eeq
The diagonal elements are the squares of the standard deviations,
$\{\tilde{\sigma}_i\}$,
of the calculated values $\{{\cal C}_i\}$ resulting from uncertainties
in the model parameter values. 
Here $C_{ij}\equiv \tilde{\sigma}_{ij}/[\tilde{\sigma}_i\tilde{\sigma}_j]$
are the correlation coefficients between the observables
${\cal C}_i$ and ${\cal C}_j$.

In principle, the best estimate
for the observables $\{{\cal C}_i\}$
is neither that resulting from using the most likely parameter values
$\{\alpha^0_k\}$ nor that calculated with the best estimate of
the model parameters, $\{\tilde{\alpha}_k\}$.
In our applications 
the distinction between the different estimates is mostly one of principle
since the different estimates yield practically identical results.
We shall generally adopt the observable values
calculated with the optimal parameter values, $\{\alpha^0_k\}$, as our estimate
while the associated uncertainties and correlations will be obtained
on the basis of the entire ensemble, as expressed in Eq.~(\ref{sigmaij}).

\section{Experimental data}
\label{data}

We discuss here the experimental data used in our study.

\subsection{Mean neutron multiplicity}

The mean number of prompt neutrons emitted following neutron-induced
fission of $^{239}\textrm{Pu}$ has been measured in a number of
experiments \cite{FrehautNu,NurpeisovNu,CondeNu,BoldemanNu,HopkinsNu}
and was reviewed by Fort {\it et al.} \cite{FortNu}.
Figure~\ref{f:nuBarFigure} shows a selection of this data 
as well as the associated ENDF/B-VII evaluation \cite{ENDFb7}. 
We employ the ENDF evaluation as an approximate average
of the experimental numbers 
and we assign a 0.5\%  uncertainty to $\nubar$.

\begin{figure}[b]	
\begin{center}
\includegraphics[width=3.1in]{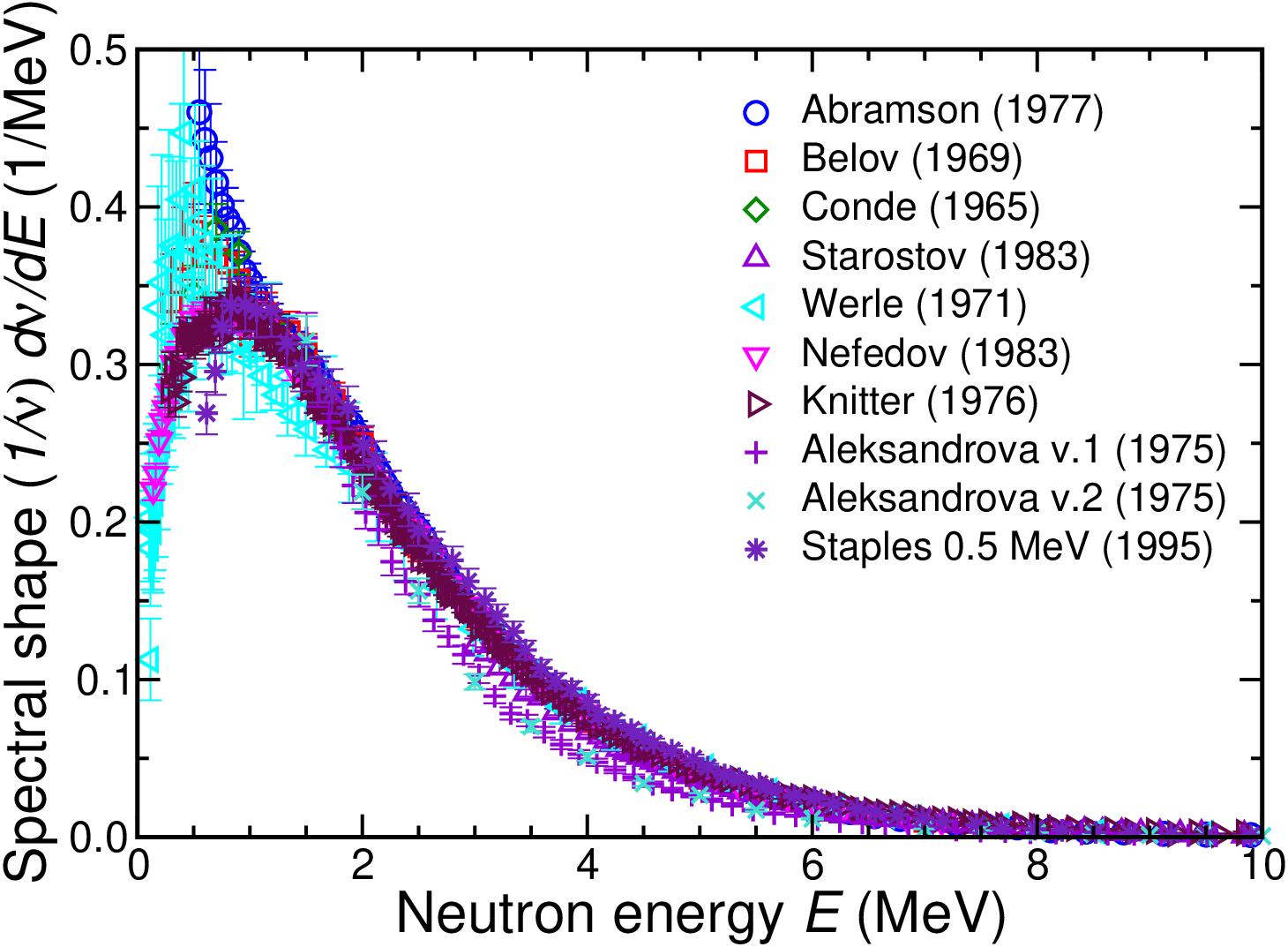}
\includegraphics[width=3.1in]{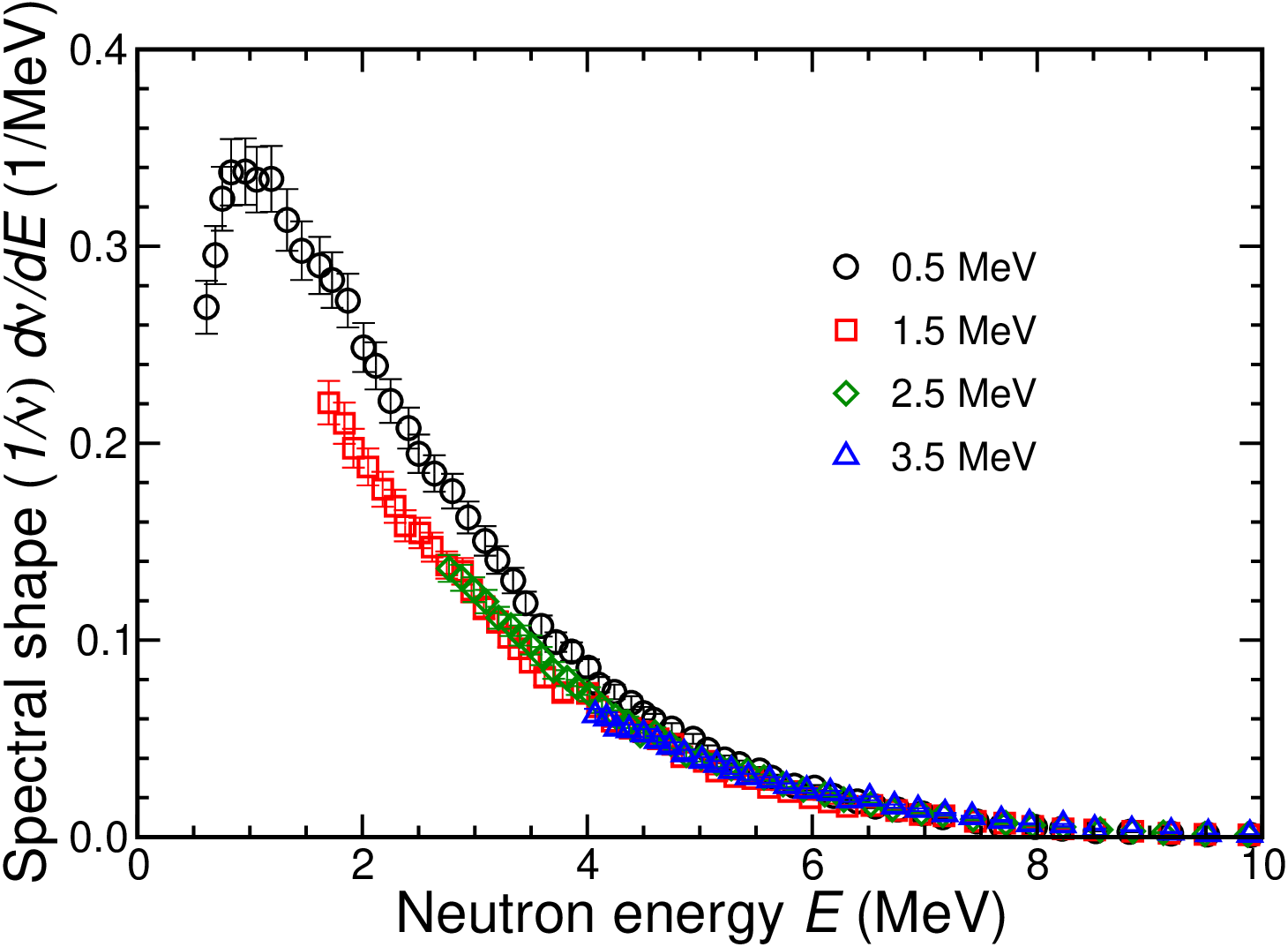}
\end{center}
\caption{(Color online)
The measured prompt neutron energy spectra, normalized to unity,
as a function of outgoing neutron energy for low incident energies from
Refs.~\protect\cite{CondePu,BelovPu,KnitterPu,AleksandrovaPu,%
AbramsonPu,NefedovPu,StaplesPu,WerlePu,StarostovPu} ({\em upper panel}) 
and for a wider range of incident energies from Ref.~\protect\cite{StaplesPu} 
({\em lower panel}).}\label{f:nspec-data}
\end{figure}		

\begin{widetext}
\begin{center}
\begin{table}[t]	
\begin{tabular}{lcclccc} \hline
Data Set & $N$~ & ~$E_{\rm min}$ (MeV)~ & ~$E_{\rm max}$ (MeV)~ & 
	~$a$ (MeV)~ & ~$b$ (MeV)~ & ~$\chi^2/N$~ \\ \hline 
Abramson  \protect\cite{AbramsonPu} & 
95 & 0.55 & 14.253 & 1.042 & 0.5294 & 2.073 \\
Aleksandrova \protect\cite{AleksandrovaPu} & 
54	& 1.503 & 11.128 & 0.914 & 0.5033 & 13.474 \\
Aleksandrova \protect\cite{AleksandrovaPu} & 
19	& 1.5 & 11 & 0.917 & 0.5033 & 14.666 \\
Belov \protect\cite{BelovPu} & 18 & 0.3 & ~7 & 0.991 & 0.5033 & 0.868 \\
Conde  \protect\cite{CondePu} & 13 & 0.3 & ~7.5 & 0.975 & 0.5365 & 1.121 \\
Knitter \protect\cite{KnitterPu} & 183 & 0.28 & 13.87 & 1.030 & 0.5040 
& 1.529 \\
Nefedov \cite{NefedovPu} & 65 & 0.139 & ~7.15 & 1.023 & 0.5053 & 0.765\\
Starostov \protect\cite{StarostovPu} & 65 & 3.007 & 11.2 & 0.995 & 0.5288 
& 3.890 \\
Werle \protect\cite{WerlePu} & 79 & 0.104 & ~9.5 & 1.035 & 0.5263 & 4.244 \\
Staples (0.5 MeV) \protect\cite{StaplesPu} & 68 & 0.615 & 16 & 1.026 & 0.5005
& 4.067 \\
Staples (1.5 MeV) \protect\cite{StaplesPu} & 59 & 1.7 & 15.2 & 1.009 & 0.5025
& 8.137 \\
Staples (2.5 MeV) \protect\cite{StaplesPu} & 51	& 2.77 &  14.4 & 1.0276 
& 0.5025 & 4.018 \\
Staples (3.5 MeV) \cite{StaplesPu} & 38	& 4.07 & 13.8 & 1.0354 & 0.5025
& 8.033 \\ \hline
\end{tabular}
\caption{For each data set is listed the number of points $N$,
the minimum and maximum measured outgoing neutron energies, 
the fitted Watt parameters $a$ and $b$,
and the associated $\chi^2$ per degree of freedom.}
\label{tabl:Watt}
\end{table}
\end{center}
\end{widetext}

\subsection{Prompt neutron spectrum}
\label{spectra}

Our statistical analysis will also incorporate
the measured prompt neutron spectrum
\cite{CondePu,BelovPu,KnitterPu,AleksandrovaPu,AbramsonPu,NefedovPu,StaplesPu}
as given in the EXFOR/CSISRS database.
Wherever the experimental uncertainties are not given
we have used an  uncertainty of 5\% 
in the calculation of $\chi^{2}$. This is likely an under-estimate of the real uncertainty.
The various data sets are shown in Fig.\ \ref{f:nspec-data}.
The bulk of the data are obtained for low incident neutron energies,
$E_n \lesssim 0.5\,{\rm MeV}$.
The remaining data have been taken 
by Staples \emph{et al}.~\cite{StaplesPu} at $E_{n}= 0.5$, 1.5, 2.5, and 3.5 MeV.

The data in the top panel of Fig.~\ref{f:nspec-data} were taken 
for incident energies below 0.5~MeV and are not absolutely normalized.
In order to compare the data sets with each other 
and with our calculated spectra, we normalize all data sets to unity
(while preserving the spectral shapes).
For this purpose, we fit the observed energy spectra to a Watt spectrum, 
\begin{equation}
\frac{dN}{dE}\, =\, N_0\, {\rm e}^{-E/a} \sinh \sqrt{2E/b}\,\, ,
\label{eq:Watt}
\end{equation}
where the normalization $N_0$ is determined by demanding that
the integral over $E$ yield unity.
Table~\ref{tabl:Watt} lists for each data set the number of data points,
the minimum and maximum neutron energies observed, 
the Watt parameters $a$ and $b$ obtained  by the fitting procedure, 
and the associated minimum $\chi^2$ per degree of freedom.
The value of $a$ is $\approx 1$~MeV within the uncertainties of the fits
for all but the Aleksandrova sets where $a\approx0.91$~MeV.  
The value of $b$ is 0.50-0.56 MeV in all cases.

The data on the neutron spectra cover a wide energy range, $0.1 <E< 14$ MeV. 
In the lowest $E$ range, $E \lesssim 0.5$ MeV, the neutron yields generally 
increase with $E$, reaching a maximum somewhere between
0.5 and 1 MeV, and decreasing again above 1 MeV.  There is significant 
disagreement between the data sets in this energy region. 
In particular, the data of Belov \emph{et al}.~\cite{BelovPu}, Werle
\emph{et al}.~\cite{WerlePu}, and Abramson \emph{et al}.~\cite{AbramsonPu}
have relatively large uncertainties and include points noticeably
higher than the remaining data. 
Curiously, the peak of the $E_n = 0.5$ MeV spectrum
from Staples {\em et al}.~\cite{StaplesPu}
is significantly narrower than those of the other data sets. 
At higher outgoing energies, $E \gtrsim 2$ MeV,
all the data sets closely follow each other, except for those from
 Aleksandrova \emph{et al}. \cite{AleksandrovaPu} which are systematically
lower.  Indeed, the Aleksandrova sets are rather poorly represented by the
Watt fits, having the largest $\chi^2$ per degree of freedom, 
see Table~\ref{tabl:Watt}.

Some of the discrepancies between the data sets may be due to the 
incompleteness of the individual sets in parts of the energy range.  
For example, the Aleksandrova \cite{AleksandrovaPu} and Starostov 
\cite{StarostovPu} sets are only available for $E$ above 1.5 and 3 MeV,
respectively, so that the Watt fits may match the high energy tail of the
spectrum but cannot represent the peak region and below.  
Similarly, sets that cover the region $E<7$ MeV may not 
give as good fits to the high-energy part of the spectrum.  
When $E_n \geq 1.5$ MeV, the minimum outgoing energy $E$ 
measured by Staples {\em et al.}\ \cite{StaplesPu}
(shown in the lower panel of Fig.\ \ref{f:nspec-data})
is always greater than $E_{n}$.  
Thus these data sets do not provide much information 
on the softer part of the spectrum and the back extrapolation
by means of the Watt form is somewhat unreliable since the measured
hard spectra do in fact not fit a Watt shape very well,
as reflected by the large values of $\chi^2$ in Table~\ref{tabl:Watt}.

\subsection{Fission fragment energies}

\begin{figure}[t]
\includegraphics[width=3.1in]{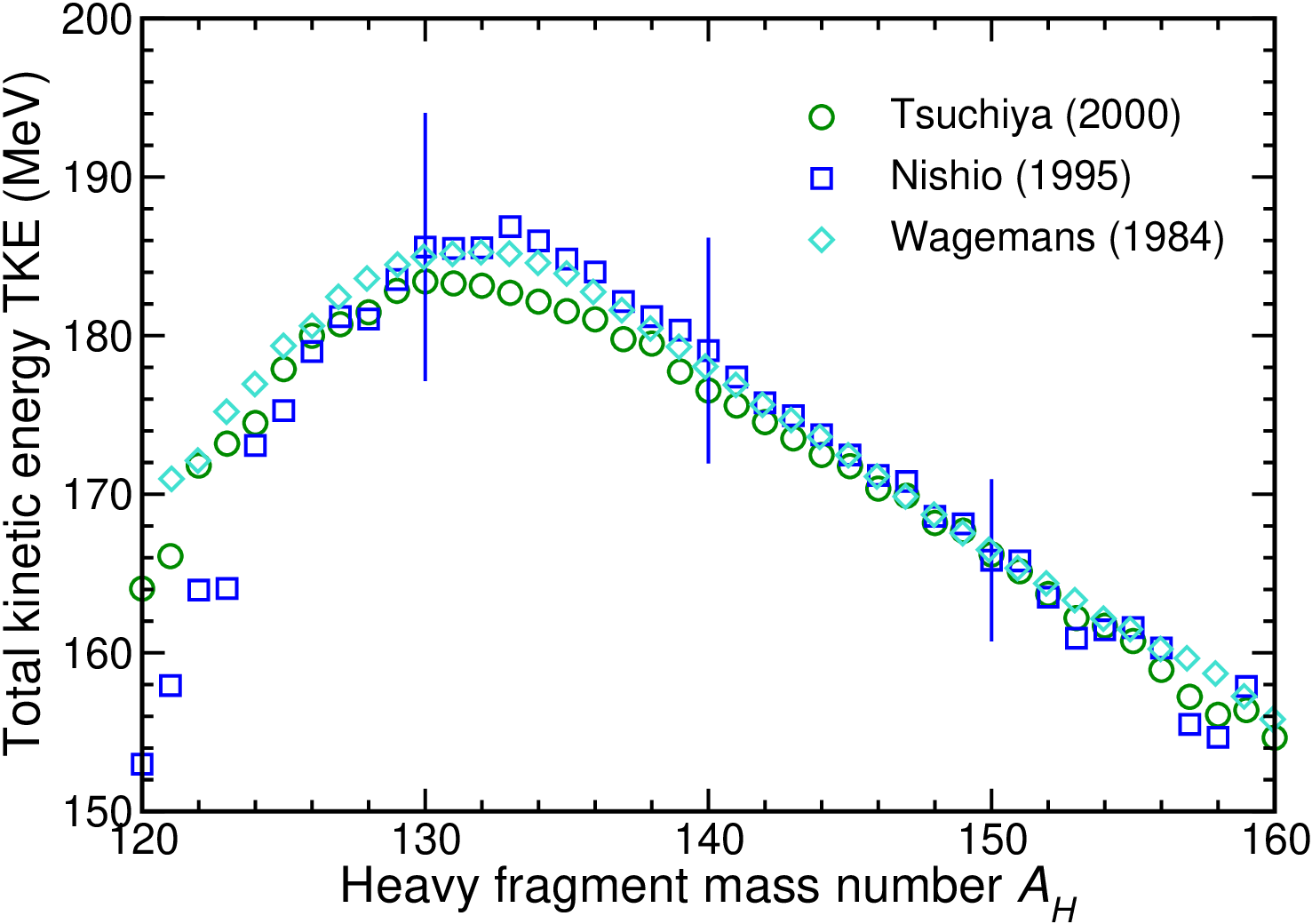}
\caption{(Color online)
The average TKE as a function of the heavy fragment mass $A_H$, 
from Refs.~\protect\cite{WagemansPu,NishioPu,TsuchiyaPu}.}
\label{cap:tke-of-ah-expt}
\end{figure}

Several measurements of the total kinetic energy (TKE) 
of the two fission fragments can be found in the literature. 
Figure~\ref{cap:tke-of-ah-expt} shows the principal measurements of 
the mean TKE as a function of the mass number of the heavy fragment, $A_H$, 
which were made by 
Wagemans \emph{et al}.~\cite{WagemansPu},
Nishio \emph{et al}.~\cite{NishioPu}, 
and Tsuchiya \emph{et al}.~\cite{TsuchiyaPu}.
(The mass number of the heavy fragment is found by simultaneously measuring 
the velocities and energies of both fragments \cite{NishioPu}.
No experimental uncertainties are given for these results, 
neither for the mass number nor for the reported TKE.)
The data exhibit a significant dip near symmetry
and fall off steadily for large asymmetries,
resulting in a maximum at $A_H\approx133$.
The different data sets generally agree well for large $A_H$
but they exhibit a significant spread near symmetry.
Furthermore, Ref.~\cite{NishioPu} also provides the full-width at half-maximum
(FWHM) of the TKE distribution at selected values of $A_H$.
These also decrease at large $A_H$,
reflecting the fact that the TKE spectrum softens,
presumably because the mutual Coulomb repulsion between the two nascent
fragments decreases with larger asymmetry.

\section{Generation of fission events}
\label{model}

We have adapted the recently developed fission model \code\ \cite{RV}
for the present purpose of calculating the neutron spectrum 
in terms of a set of well-defined model parameters.
Since this is the first practical application of \code,
we describe its main physics ingredients below.

The code follows the temporal sequence of individual fission events
from the initial excited fissionable nucleus, 
$^{240}{\rm Pu}^*$ in the present case,
through a scission configuration of the two nascent fragments,
to the subsequent neutron evaporation from the fully accelerated fragments.
The competition between fission
and neutron emission from the fissioning nucleus ($2^{\rm nd}$ chance fission)
has not yet been implemented in the code.
Consequently, we restrict our discussion to energies below
 $5.5$~MeV.

\subsection{Fission mass and charge partition}
\label{split}

The fission process is initiated when a neutron with a specified initial 
energy $E_n$ is absorbed by a fissile nucleus to form a compound nucleus $^AZ$
with a certain excitation energy.  
The compound nucleus subsequently splits into a heavy fragment $^{A_H}Z_H$ 
and a complementary light fragment $^{A_L}Z_L$. 
In its present early form, \code\ selects the mass and charge partitions
on the basis of existing experimental data.
For the present study, 
we use fits to the thermal and fast $^{239}{\rm Pu}(n,f)$
fission product mass yields measured by England and Rider \cite{EnglandRider}
in combination with the charge distributions obtained by 
Reisdorf {\em et al}.~\cite{ReisdorfNPA177}.

The fits assume that the mass product yields $Y(A_p)$ 
exhibit three distinct fission modes that can be represented
in terms of suitable gaussians,
\begin{equation}\label{eq:defs1s2}
Y(A_p)\ =\ S_1(A_p)+S_2(A_p)+S_L(A_p)\ .
\end{equation}
The first two terms result from asymmetric fission modes
associated with the spherical shell closure at $N=82$
and the deformed shell closure at $N=88$ respectively,
while the last term results from a symmetric, so-called super-long, mode
which is relatively insignificant \cite{Brosa}.
The specific forms of these terms are
\begin{equation}
S_i = {N_i\over\sqrt{2\pi}\sigma_i}\left[
 {\rm e}^{-(A-\bar{A}-D_i)^2/2\sigma_i^2}
+{\rm e}^{-(A-\bar{A}+D_i)^2/2\sigma_i^2}
\right]
\end{equation}
for $i=1,2$ and
\begin{equation}
S_L = {N_L\over\sqrt{2\pi}\sigma_L}\,{\rm e}^{-(A-\bar{A})^2/2\sigma_L^2}\ .
\end{equation}
Here $\bar{A}=\mbox{$1\over2$}(A_0-\bar{\nu})$, 
where $A_0=240$ is the mass number of the fissioning nucleus
and $\nubar$ is the average total multiplicity of evaporated neutrons.
(While there exist more detailed data for \emph{e.g.} $^{235}\textrm{U}(n,f)$
that give the yields as a function of both mass and total kinetic energy, 
$Y\left(A_p,{\rm TKE}\right)$, for several values of $E_{n}$ \cite{Hambsch}, 
such data are not yet available for Pu.)

\begin{table}[hbp]
\begin{tabular}{ccc}
\hline 
~~Parameter~~&
~~Thermal~~&
~~~~~Fast~~~~~\tabularnewline
\hline
$\bar{A}$&
118.5&
117.5\tabularnewline
\hline 
$N_{1}$&
0.7574&
0.7355\tabularnewline
$D_{1}$&
20.81&
20.96\tabularnewline
$\sigma_{1}$&
5.626&
5.711\tabularnewline
\hline 
$N_{2}$&
0.2417&
0.2623\tabularnewline
$D_{2}$&
14.95&
15.14\tabularnewline
$\sigma_{2}$&
2.546&
2.627\tabularnewline
\hline 
$N_{L}$& 0.0018& 0.0044\tabularnewline
$\sigma_{L}$&
1.824&
2.511\tabularnewline
\hline
\end{tabular}
\caption{The fit parameters of the three fission modes
for thermal and fast neutron-induced fission on $^{239}\textrm{Pu}$.}
\label{cap:yofatab}
\end{table}

The values of the parameters in the fits to $Y(A_p)$ are given
in Table \ref{cap:yofatab} for either thermal or fast fission.
The normalization is chosen such that $\sum_AY(A)=2$
since each event leads to two products.
Consequently we have $2N_1+2N_2+N_L=2$,
apart from a negligible correction 
because $A_p$ is discrete quantity bounded both from below and above.
It should be noted that the symmetric component contributes
only 1-2 per mille of the yield.

While these fits are to the fission \emph{product} yields,
the \code\ simulation requires fission \emph{fragment} yields,
 \emph{i.e.} the probability distribution for obtaining a given mass partition 
at scission, before neutron evaporation has begun,
We take $\bar{A}\approx\mbox{$1\over2$}A_0$,
but keep the displacements $D_i$ and the widths $\sigma_i$ unchanged.
We use the thermal fits for $E_{n} < 1$ MeV and
the fast fits for $1 <E_{n} < 5.5$ MeV, the
highest value of $E_{n}$ considered here. The
change in the fit parameter values with incident neutron energy should,
of course, be more continuous than we have implemented here but the
change is most important in areas where the yields are low: the tails
of the gaussians where fission is most asymmetric and in the case
of symmetric fission. At even higher energies, symmetric fission
(the $S_{L}$ component) grows increasingly important, 
filling in the dip at symmetry.
The width $\sigma_{2}$ also increases, broadening the asymmetric tails.

\begin{figure}[t]
\includegraphics[width=3.1in]{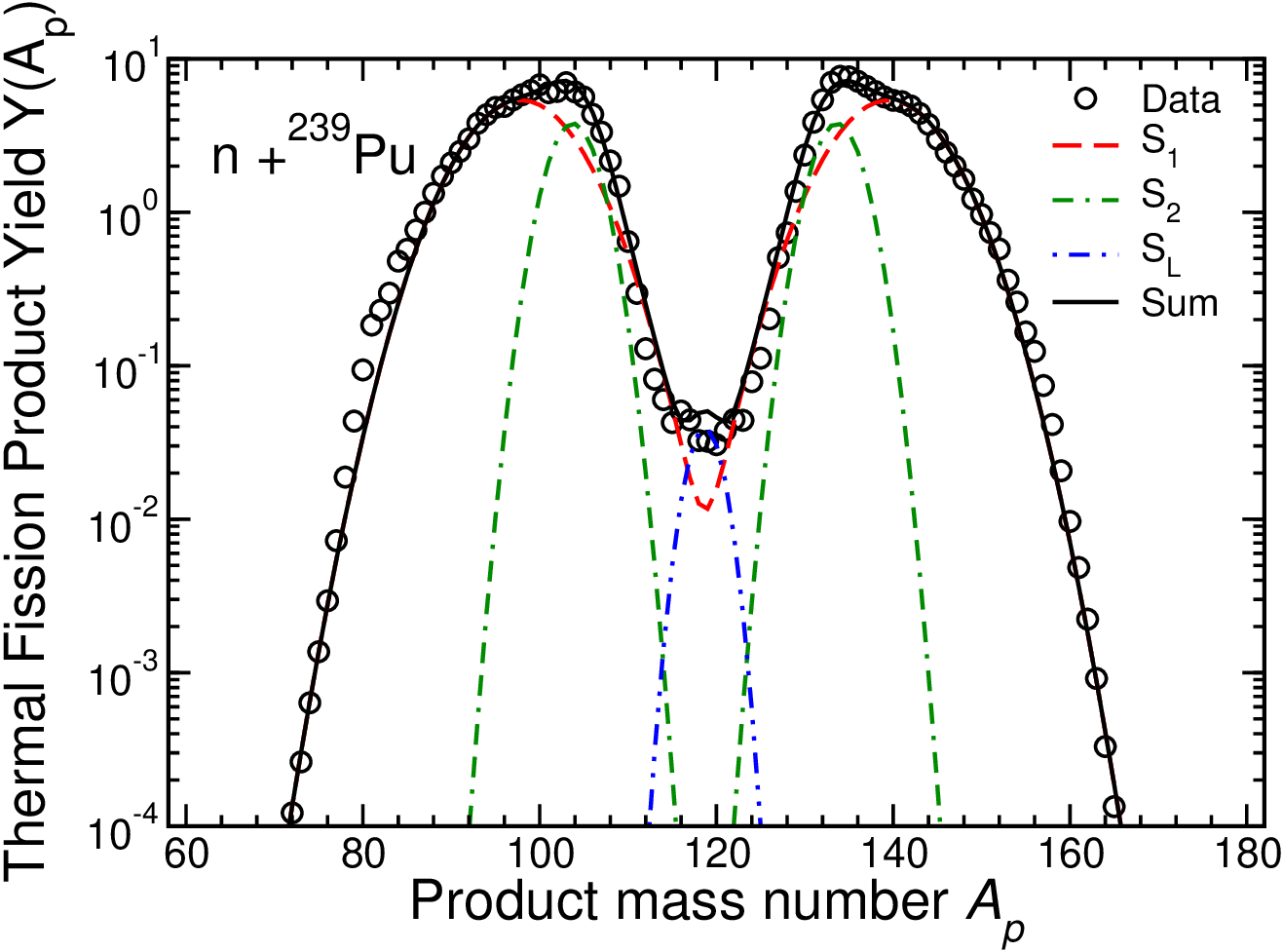}
\caption{(Color online)
The fission product yield as a function of
fragment mass for thermal fission. The data are from 
Ref.~\protect\cite{EnglandRider}
while the curves are a five-gaussian fit to the data.}\label{cap:yofathermal}
\end{figure}

\begin{figure}[t]
\includegraphics[width=3.1in]{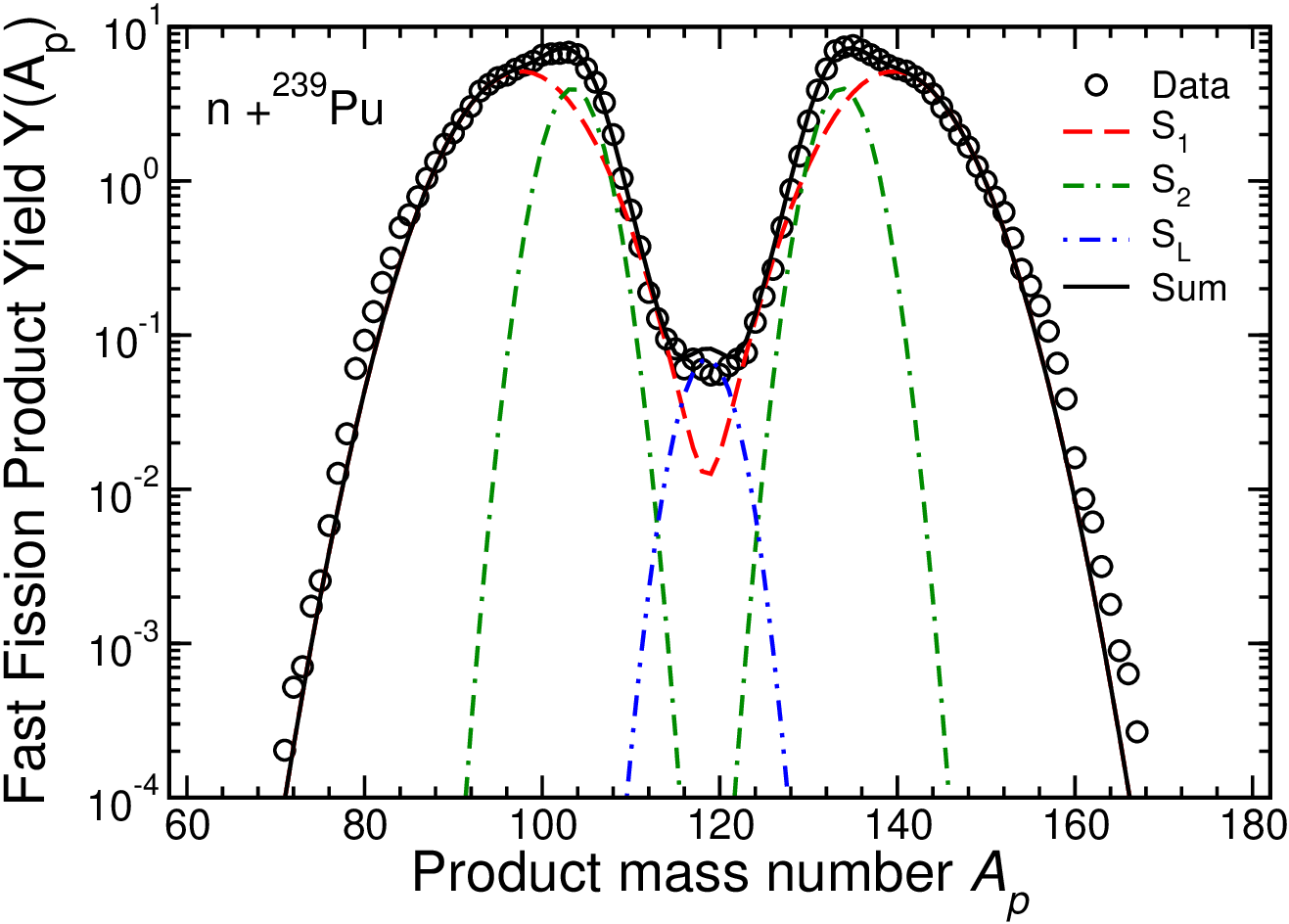}
\caption{(Color online)
Same as Fig.~\protect\ref{cap:yofathermal}, 
but for fast-neutron induced fission.}\label{cap:yofafast}
\end{figure}

The resulting fits are compared to the data in Figs.~\ref{cap:yofathermal}
and \ref{cap:yofafast}. The agreement with the tabulated percentage
yields is quite good, especially in the regions where the yields are
highest and which thus contribute the greatest number of events. Equation
(\ref{eq:defs1s2}) does not perfectly describe the tails at high
and low fragment mass. We have also tried a fit with 5 independent
gaussians, \emph{e.g.} allowing $N_{i}$, $D_{i}$ and $\sigma_{i}$
to vary independently on the low and high sides of $\overline A$, and found
that the fit does not significantly improve as a result. We note also
that the width of the $S_{L}$ component is not as large as found
in other actinides where the yields have been decomposed in a similar
fashion \cite{Hambsch}.

Once the gaussian fit has been fixed,
it is straightforward to make a statistical selection of the fragment
mass number $A_f$.  The mass number of the partner fragment is then readily
determined since we assume $A_L+A_H=A_0$.

The fragment charge, $Z_f$, is selected subsequently.
For this we follow Ref.~\cite{ReisdorfNPA177} and employ a gaussian form,
\begin{equation}
P_{A_f}(Z_f)\ \propto\ {\rm e}^{-(Z_f-\bar{Z}_f(A_f))^2/2\sigma_Z^2}\ ,
\end{equation}
with the condition that $|Z_f-\bar{Z}_f(A_f)|\leq5\sigma_Z$.
The centroid is determined by requiring that the fragments have, on average,
the same charge-to-mass ratio as the fissioning nucleus,
$\bar{Z}_f(A_f)=A_fZ_0/A_0$.  The dispersion is 
the measured value, $\sigma_Z=0.5$ \cite{ReisdorfNPA177}.
The charge of the complementary fragment then follows using $Z_L+Z_H=Z_0$.

\subsection{Fragment energies}

Once the partition of the total mass and charge among the two fragments
has been determined, 
the $Q$ value associated with that particular channel follows as the 
difference between the mass of the excited compound nucleus, $^{240}{\rm Pu}^*$,
and ground-state masses of the two fragments,
\begin{equation}
Q_{LH}\ =\ M(^{240}{\rm Pu}^*) - M_L - M_H\ .
\end{equation}
\code\ takes the required nuclear ground-state masses
from the compilation by Audi {\em et al.}~\cite{Audi},
supplemented by the calculated masses of M{\"o}ller {\em et al.}~\cite{MNMS}
where no data are available.
The $Q_{LH}$ value for the selected fission channel is then divided up between
the total kinetic energy (TKE) and the total excitation energy (TXE)
of the two fragments.
The specific procedure employed is described below. 

First, the average value of TKE is determined on the basis of 
the Coulomb potential between the two fragments at scission,
\begin{equation}
\overline{\rm TKE}\ =\ e^2\, \frac{Z_L Z_H}{c_L + c_H + d_{LH}}\ .
\label{TKEdef}
\end{equation}
In the scission configuration, the two nascent fragments are
assumed to have spheroidal shapes and be positioned coaxially 
with a tip separation of $d_{LH}$.
The associated major axes are 
$c_i=r_0 A_i^{1/3}/[1-\mbox{$2\over3$}\varepsilon(Z_i,A_i)]$ 
with $r_0 = 1.2$ fm.
We use the values for the spheroidal deformation parameter 
$\varepsilon(Z_i,A_i)$
 calculated in Ref.~\cite{MNMS} which include shell effects.
The denominator of Eq.~(\ref{TKEdef}) is thus the distance between 
the centers of the two fragments and the above expression represents
the monopole-monopole term of the mutual Coulomb interaction energy.

The tip separations $\{d_{LH}\}$ are important parameters in the model
since they determine the (average) fragment kinetic energies
and hence, by energy conservation, also the total fragment excitation  
that is available for neutron emission. 
Thus the neutron emission is quite sensitive to the specified values
of $\{d_{LH}\}$ and they deserve careful consideration. 
Furthermore, since the TKE is closely related to the Coulomb potential at
scission, these parameters contain valuable information 
about the scission configurations.

\begin{figure}[t]
\includegraphics[width=3.1in]{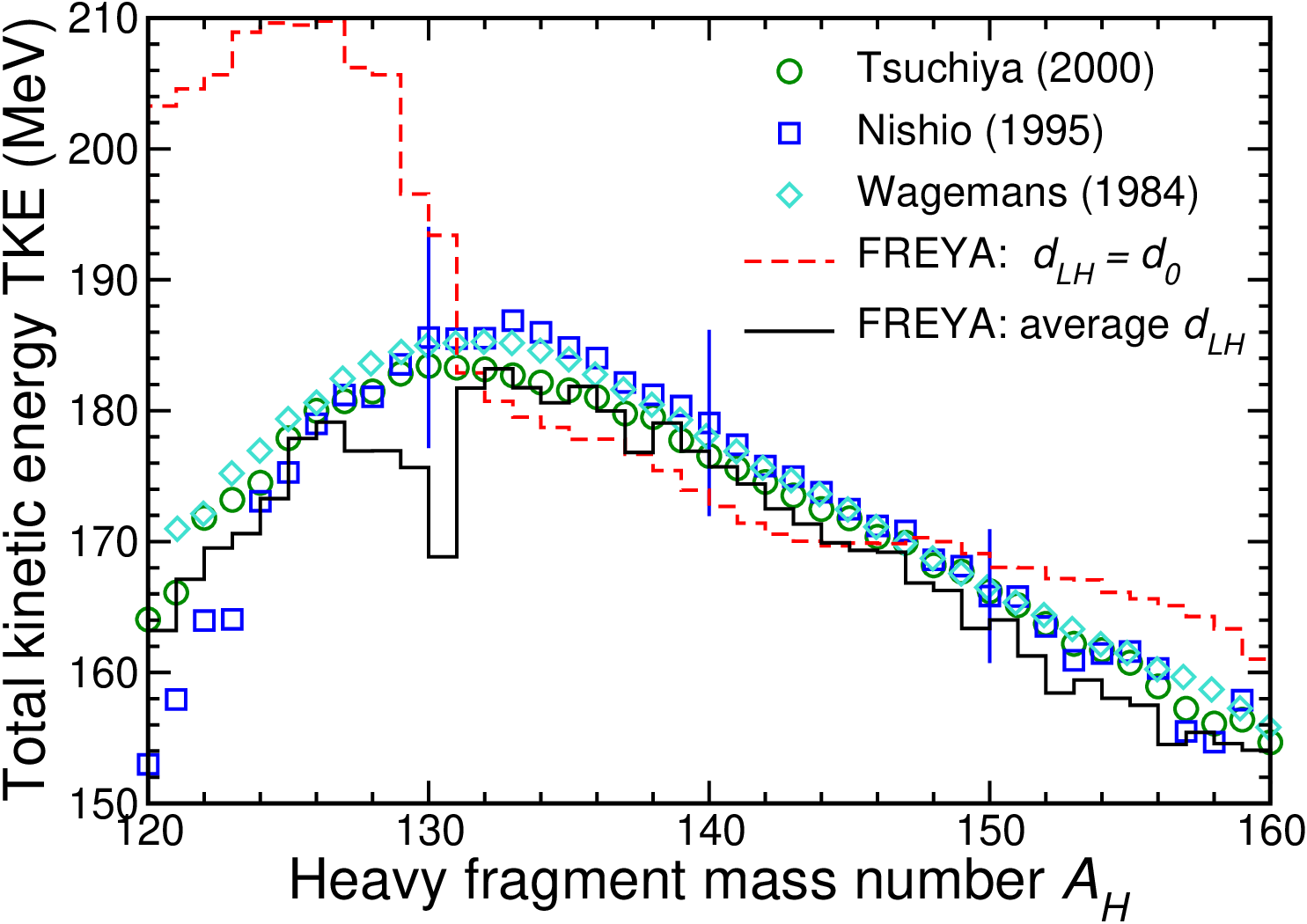}
  \caption{(Color online)
The measured average TKE as a function of the mass number of
the heavy fragment \protect\cite{WagemansPu,NishioPu,TsuchiyaPu} 
compared to \code\ 
calculations with a constant tip separation of $d_0$=4.05 fm and the 
average distance extracted from Fig.~\protect\ref{variabled}.}
\label{tkefixedvsvard}
\end{figure}

Figure~\ref{tkefixedvsvard} shows the mean total fragment kinetic energy 
as a function of mass number of the heavy fragment
as obtained by using a common tip separation $d_0$ for all fission channels.
A comparison to the data \cite{WagemansPu,NishioPu,TsuchiyaPu} shows
significant discrepancies near symmetry where the calculated TKE
exhibit an enhancement whereas the data have a dip.

\begin{figure}[t]
\includegraphics[width=3.1in]{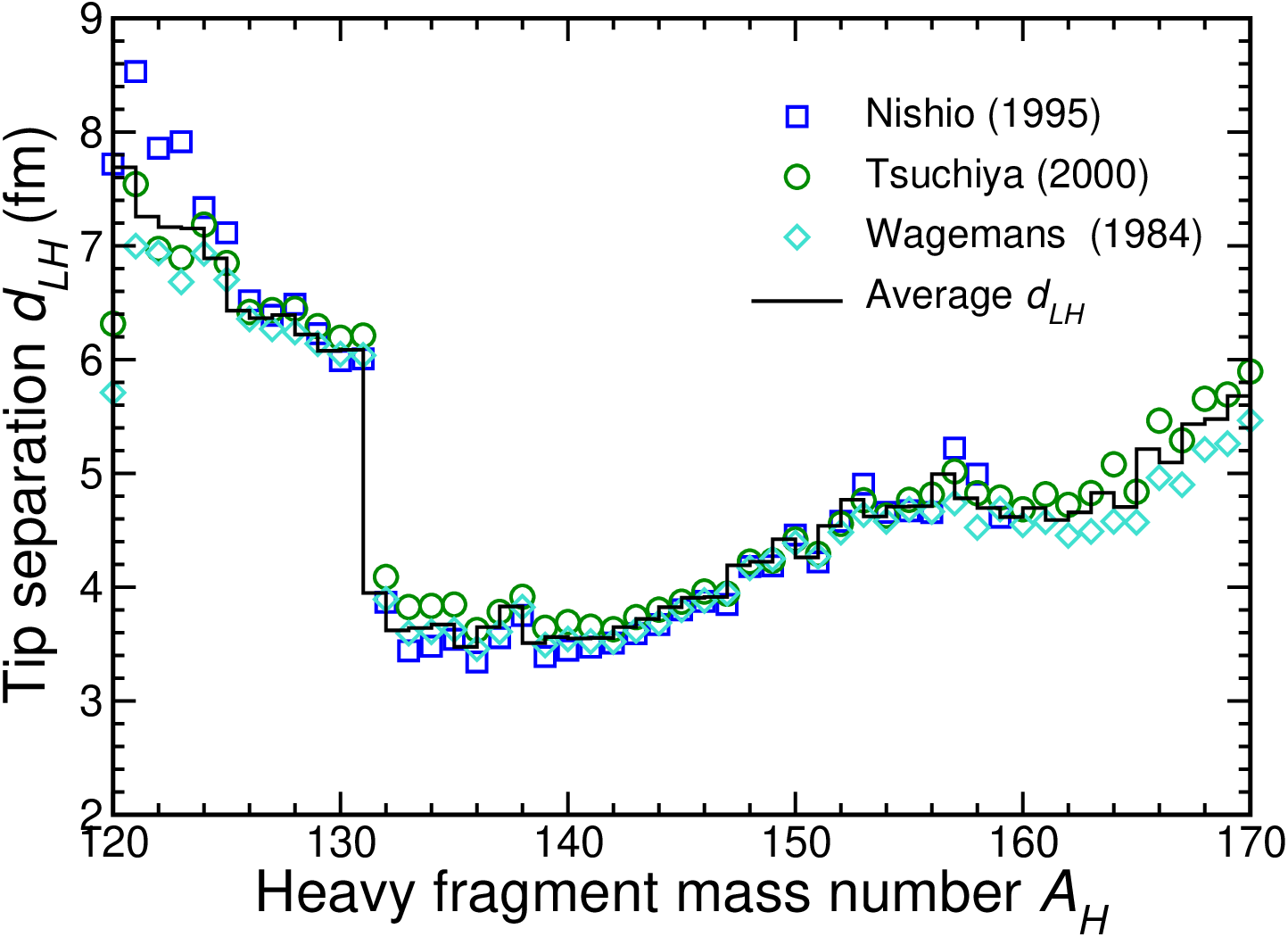}
  \caption{(Color online)
The tip separation $d_{LH}$ fitted to the TKE values 
measured at thermal energies \protect\cite{WagemansPu,NishioPu,TsuchiyaPu},
with the deformation radii extracted from the mass model of 
Ref.~\protect\cite{MNMS}.}
\label{variabled}
\end{figure}

To account for the dependence of the tip separation on the mass partition,
we took the average of the data sets shown in Fig.~\ref{tkefixedvsvard} 
and extracted the average tip separations $\underline{d}_{LH}$ 
shown in Fig.~\ref{variabled},
assuming that the two fragments have the same charge-to-mass ratio.
Near symmetric fission, $\underline{d}_{LH}$ is large, 7-8 fm at $A_H = 120$, 
with a steep drop to less than 4 fm for $A_H \geq 132$.  
Near symmetry, the plutonium fission fragments are mid-shell nuclei
subject to strong deformation.  Thus the scission configuration will contain
significant deformation energy and a correspondingly large distance between
centers, resulting in low TKE.  At $A_H = 132$, the heavy fragment is close
to the doubly-magic closed shell with $Z_H = 50$, $N_H = 82$ and 
is resistant to distortions away from its spherical shape.  
However, the complementary light fragment is far from a closed shell 
and is significantly deformed, having thus a large value of $c_L$
which then results in a small tip separation $\underline{d}_{LH}$ 
and a large TKE.  
The passage of the heavy fragment mass through the doubly-magic region 
results in the dip in calculated TKE around $A_H \sim 130$, 
see Fig.~\ref{tkefixedvsvard}.

The TKE values shown in Fig.~\ref{tkefixedvsvard} were obtained in 
experiments with incident neutrons of very low energy and there are no other
higher-energy data to show how TKE$(A_H)$ evolves with incident neutron energy.
At each higher incident energy $E_n>E_{\rm thermal}$,
we use tip separations obtained by scaling 
those fitted at thermal energies, 
\beq 
d_{LH}(E_n)=s(E_n)d_{LH}(E_{\rm thermal}) \, \, , \label{sparam}
\eeq
and use the common scaling factor $s(E_n)$ 
as one of the adjustable model parameters in our fits to the neutron spectra.
The average neutron multiplicity is very sensitive to this scale factor which,
as we shall show, is greater than but very close to unity for the entire
energy range studied.

\begin{figure}[t]
\includegraphics[width=3.1in]{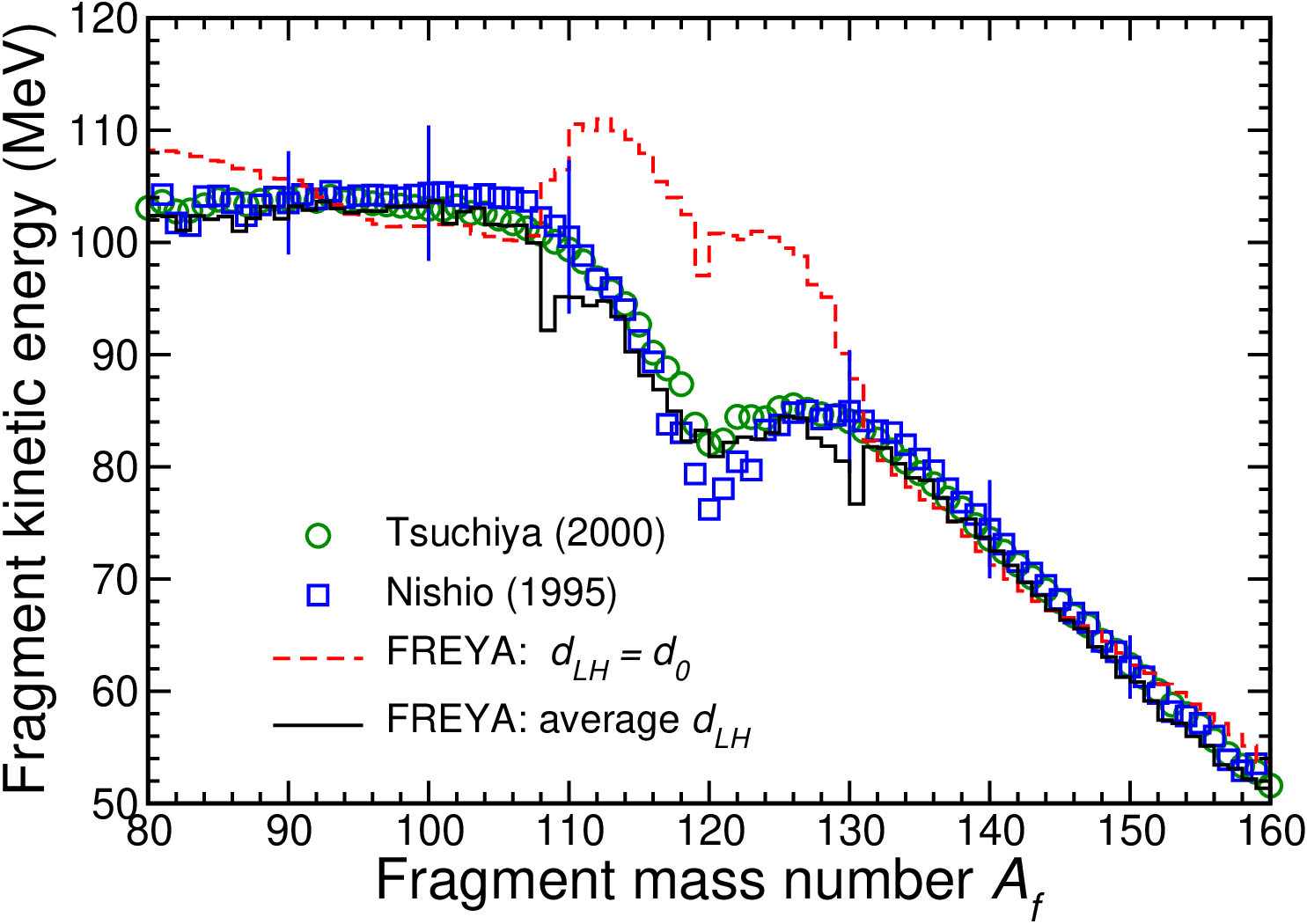}
  \caption{(Color online)
The average fragment kinetic energy
as a function of fragment mass from Refs.~\protect\cite{NishioPu,TsuchiyaPu}
compared to \code\ 
calculations with a constant tip separation of $d_0=4.05$~fm 
and the average distance extracted from Fig.~\protect\ref{variabled}.}
\label{skefixedvsvard}
\end{figure}

\begin{figure}[t]
\includegraphics[width=3.1in]{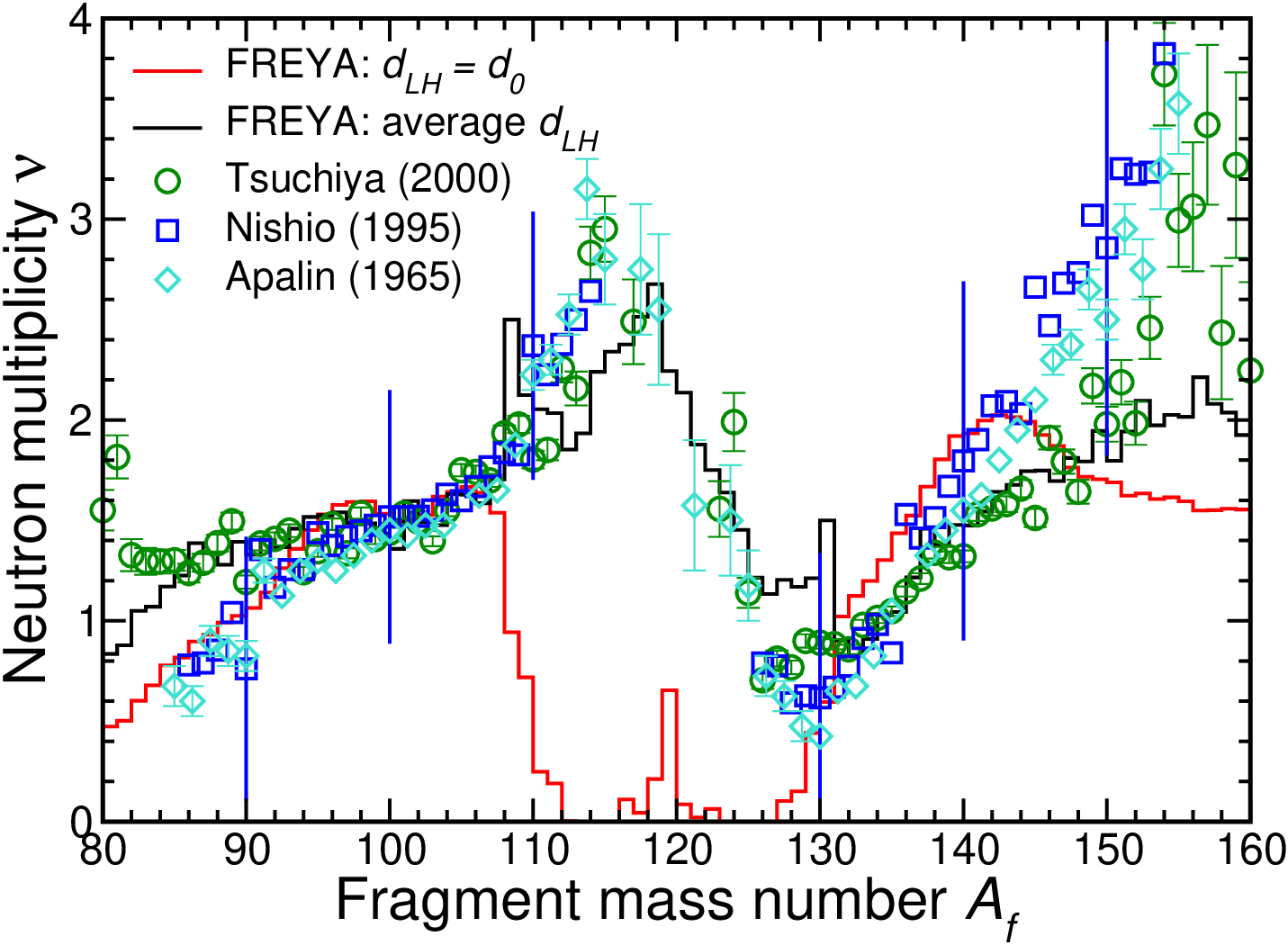}
  \caption{(Color online)
The average neutron multiplicity as a function of the fragment
mass from Refs.~\protect\cite{NishioPu,TsuchiyaPu,ApalinPu} compared to \code\ 
calculations with a constant tip separation of $d_0=4.05$~fm 
and the average distance extracted from Fig.~\protect\ref{variabled}.}
\label{nuofafixedvsvard}
\end{figure}

As shown in Fig.~\ref{tkefixedvsvard}, 
the scaled tip separations lead to a very good agreement with the TKE data.  
With this means of fixing $d_{LH}$, the
TKE is no longer overestimated near symmetry, leading to a better approximation
of the individual fragment kinetic energy as well as the neutron multiplicity 
as a function of fragment mass, overestimated and underestimated respectively
with a constant value of $d_{LH}$, as shown in Figs.~\ref{skefixedvsvard} and
\ref{nuofafixedvsvard}.  The variable
$d_{LH}$ also correctly produces the dip in the single fragment kinetic energy
shown in Fig.~\ref{skefixedvsvard}.  The small dips in the fragment kinetic
energy at $A = 110$ and 130 correspond to the dip at $A_H \sim 130$ in
Fig.~\ref{tkefixedvsvard}. 

The overestimate of the total fragment kinetic energy with a constant $d_{LH}$ 
leaves insufficient excitation energy available for neutron
evaporation near symmetry, resulting in the near absence of neutron
emission in Fig.~\ref{nuofafixedvsvard} in this case.
On the other hand, with $d_{LH}$ from Fig.~\ref{variabled}, there is a
peak in the neutron emission near symmetry, followed by a drop for $A > 120$,
resulting in the characteristic sawtooth shape of $\nu (A)$.   
The decrease in KE for these values of $A$ gives
small peaks in the neutron multiplicity at the same values of $A$.
Interestingly enough, the calculations with both fixed and variable $d_{LH}$,
give the same $\overline\nu$ even though $\overline{\nu}(A_f)$ 
is very different in the two cases.  
It is easy to see why this is true by looking at 
Figs.~\ref{cap:yofathermal} and \ref{nuofafixedvsvard} together.  
Symmetric fission does not contribute significantly to the total yield, 
$Y(A_f)$.  Most of
the fragment yield is around $A_L\sim 100$, $A_H \sim 140$.  
The variable $d_{LH}$ gives more neutrons for symmetric fission 
and in regions of high $A_H$ (low $A_L$) 
with lower yields and fewer neutrons where $Y(A_f)$ is large to obtain the
same $\overline \nu$ as the constant $d_{LH}$ where the neutrons from symmetric
fission are effectively absent.

Once the average total fragment kinetic energy has been determined, 
the average combined excitation energy in the two fragments follows
automatically by energy conservation,
\begin{equation}
Q_{LH} - \overline{\rm TKE}\ =\ \overline{\rm TXE}\
=\ \overline{E}_L^*+\overline{E}_H^*\ .
\end{equation}
The last relation indicates that the total excitation energy
is partitioned between the two fragments.
The variation of the total mean excitation energy with fragment mass
is similar to that of $\nubar(A)$ in Fig.~\ref{nuofafixedvsvard}.

\code\ assumes that the excitation energy is partitioned statistically, 
as it would be if the two fragments were in mutual thermal equilibrium.  
Consequently, 
$\overline{\rm TXE}$ is divided in proportion to the heat capacities
of the nascent fragments,
\begin{equation}
\overline{E}_i^*\ =\ 
{\tilde{a}_i \over \tilde{a}_L+\tilde{a}_H}\,\overline{\rm TXE}\ ,
\end{equation}
where $\tilde{a}_i$ is the level-density parameter for fragment $i$.
To take account of the microscopic structure of the individual fragments 
as well as any possible energy dependence, 
\code\ uses the functional form due to Kawano {\em et al}.\ \cite{Kawano},
\begin{eqnarray}
\tilde a_i(E_i^*) = {A_i\over e_0}
\bigg[ 1 + \frac{\delta W_i}{U_i} [1 - e^{-\gamma U_i}] \bigg]
\label{aleveldef}
\end{eqnarray}
where $U_i = E_i^* - \Delta_i$ and $\gamma = 0.05$ \cite{LemairePRC72}.  
The pairing energy of the fragment, $\Delta_i$,
and its shell correction, $\delta W_i$
are tabulated in Ref.~\cite{Kawano} 
based on the mass formula of  Koura {\em et al.}~\cite{Koura}.
Although \code\ uses the default value $e_0=7.25$\,MeV \cite{RV},
we wish to make this value adjustable, taking 
\beq 
e_0=(7.25\,{\rm MeV}) \, a \label{aparam}
\eeq
and treating the common factor $a$ as a model parameter.
We note that if the shell corrections are negligible, $\delta W\approx0$,
then this renormalization is immaterial
and the excitation energy will be shared according to mass,
$\bar{E}_i^*\propto A_i$.

The relationship between excitation energy $E_i^*$
and the temperature $T_i$ is given by
\begin{equation}
E^*_i\ =\ \tilde{a}_i T_i^2
\end{equation}
so that when the total excitation energy is shared according to the
level-density parameters $\tilde{a}_i$ then the two fragment temperatures
are equal, $T_L=T_H$.

While the equal temperature assumption is a reasonably good first 
approximation, it may be inadequate for obtaining a detailed description of
prompt neutron emission.  
Therefore we redistribute the excitation energies of the fragments,
\begin{eqnarray}
\tilde{E}^*_L = x \bar{E}^*_L\ ,\ 
\tilde{E}^*_H = \overline{\rm TKE}-\tilde{E}^*_L\ ,
\label{eeshift}
\end{eqnarray}
and treat $x$ as an adjustable model parameter.
The data indicate that the light fragments acquire more than their
``fair share'' of the energy, thus we expect that 
our statistical analysis will favor $x>1$.

After the mean excitation energies have been assigned,
\code\ considers the effect of thermal fluctuations 
in  the partitioning of the excitation energy.
For this task, \code\ assumes that the fluctuation in the excitation energy
of a nucleus is $\sigma_E^2=2\bar{E}^*T$
where $T$ is its temperature and $\bar{E}^*=\tilde{a}T^2$ its mean excitation.
Therefore, for each of the two fragments,
we sample a thermal energy fluctuation $\delta E_i^*$ 
from a gaussian distribution of variance $\sigma_i^2=2\tilde{E}_i^*T_i$
and modify the fragment excitations accordingly,
\beq
E_i^*\ =\ \tilde{E}_i^*+\delta E_i^*\ ,\ i=L, H\ .
\eeq
Due to energy conservation, there is a compensating opposite fluctuation in
the total kinetic energy, so
\begin{eqnarray}\label{tkefinal}
{\rm TKE}\ =\ \overline{\rm TKE} - \delta E^*_L - \delta E^*_H \ .
\end{eqnarray}

With both the excitations and the kinetic energies of the two fragments
fully determined, it is an elementary matter to calculate the magnitude
of their momenta and thus sample the velocities
with which they emerge after having been fully accelerated 
by their mutual Coulomb repulsion \cite{RV}.

\subsection{Neutron evaporation}

The primary fission fragments are typically sufficiently excited 
to permit the emission of one or more neutrons.
For each of the two fragments, neutron emission is treated 
by iterating neutron evaporation from each fragment.

At each step in the evaporation chain,
the excited mother nucleus $^{A_i}Z_i$
has a total mass equal to its ground-state mass
plus its excitation energy, $M_i^* = M_i^{\rm gs}+ E_i^*$.
The $Q$-value for neutron emission from the fragment
is then $Q_{\rm n}=M_i^* -M_f - m_{\rm n}$,
where $M_f$ is the ground-state mass of the daughter nucleus
and $m_{\rm n}$ is the mass of the neutron
(for neutron emission we have $A_f=A_i-1$ and $Z_f=Z_i$).
The $Q$-value is equal to the maximum possible excitation energy
of the daughter nucleus,
which occurs if the final relative kinetic energy vanishes.
The temperature in the daughter fragment is then maximal.
Thus, once $Q_{\rm n}$ is known, one may sample the kinetic energy 
of the evaporated neutron.
\code\ assumes that the angular distribution is isotropic 
in the rest frame of the mother nucleus 
and uses a standard spectral shape \cite{Weisskopf},
\beq
f_{\rm n}(E)\ \equiv\ {1\over\nubar}{d\nubar\over dE}\
\propto\  E\,{\rm e}^{-E/T_f^{\rm max}}\ ,
\eeq
which can be sampled very fast \cite{RV}.

Although relativistic effects are very small, we take them into account
in order to ensure exact conservation of energy and momentum,
which is convenient for code verification purposes.
We therefore take the sampled energy $E$ to represent the
{\em total} kinetic energy in the rest frame of the mother nucleus,
{\em i.e.}\ it is the kinetic energy of the emitted neutron
{\em plus} the recoil energy of the residual daughter nucleus.
The excitation energy in the daughter nucleus is then given by
\beq
E_d^*\ =\ Q_{\rm n}-E \,\, .
\eeq
The mass of the daughter nucleus is thus $M_d^* = M_d^{\rm 
gs} + E_d^*$.  It is possible to calculate the magnitude of the momenta 
of the two final bodies: the excited daughter and the emitted neutron.
Sampling the direction of their relative motion isotropically,
we thus obtain the two final momenta
which are subsequently boosted into the overall reference frame
by the appropriate Lorentz transformation.

This procedure repeated until 
no further neutron emission is energetically possible, when $E_d^*<S_{\rm n}$,
where $S_{\rm n}$ is the neutron separation energy for the daughter nucleus,
$S_{\rm n}=M(^{A_d}Z_d)-M(^{A_d-1}Z_d)-m_{\rm n}$.

\section{Results}
\label{results}

We now proceed to discuss our analysis.
We first describe the computational approach
and then explain how the model parameters are determined.
The resulting prompt neutron spectrum is then discussed in detail.
Finally, we present some additional observables of particular relevance.

\subsection{Computational approach}
\label{sub:Statistical-analysis-of}

\code\ is used to generate a large sample of fission events
(typically one million events for each parameter set).
For each set $m$ of such randomly selected model parameter values,
$\{s^{(m)},a^{(m)},x^{(m)}\}$,
the prompt fission neutron spectrum and $\overline\nu$ in each event $m$
are then compared to the available experimental data at the given incident
neutron energy, $E_n$. 
This allows us to assign the likelihood for that particular set
(see Sec.~\ref{method})
based on either the $\chi_m^2$ for comparison with $\nubar$ only, 
$\chi_\nubar^2$,
or on the total $\chi_m^2$ characterizing the comparison with both
$\nubar$ and the spectral shape $f_n(E)=\nubar^{-1}d\nubar/dE$,
$\chi_\nubar^2+\chi_{\rm spectra}^2$,
\begin{equation}\label{w}
w_m\ =\ 
w\{s^{(m)},a^{(m)},x^{(m)}\}\ =\ {\rm e}^{-\chi_m^{2}/2}\ .
\end{equation}
Since the weight $w_m$ depends exponentially on $\chi_m^2$,
the likelihood tends to be strongly peaked around the favored set.
It is important that the parameter sample be sufficiently dense
in the peak region to ensure that many sets have non-negligible weights.
We typically sample 2000 different parameter sets
but have verified that the results remain unchanged 
when a five times larger sample is explored.

Using this method, we can obtain those values of $s$, $a$ and $x$ 
that minimize either $\chi_\nubar^2$ or $\chi_\nubar^2+\chi_{\rm spectra}^2$.
We denote the optimal set by $\{s^0,a^0,x^0\}$.
We also obtain the corresponding correlation matrix,
as described in Sec.~\ref{method}.

\begin{widetext}
\begin{center}
\begin{table}[hpb]
\begin{center}
\begin{tabular}{ccccccc} \hline
$E_n$ (MeV) & $s^0$ & $a^0$ & $x^0$ & $\overline \nu$ 
& $\chi^2_{\overline \nu}$ & $\chi^2_{\rm spectra}/N$ \\ \hline
0.5 & ~$1.05449 \pm 0.00567$~~& 	
	~$1.10562 \pm 0.07987$~~& ~~$1.10264 \pm 0.05909$~~
	& $2.948 \pm 0.015$~& $4.26 \times 10^{-3}$ & 28.99 \\
1.5 & $1.05887 \pm 0.00585$ & 	
	$1.10426 \pm 0.07854$ & $1.10178 \pm 0.05736$
	& $3.090 \pm 0.015$ & $8.46 \times 10^{-4}$ &   9.81 \\
2.5 & $1.06590 \pm 0.00858$ & 	
	$1.10243 \pm 0.07972$ & $1.09969 \pm 0.11359$
	& $3.242 \pm 0.016$ & $1.88 \times 10^{-2}$ &   3.40 \\
3.5 & $1.06886 \pm 0.00902$ & 	
	$1.10440 \pm 0.07903$ & $1.09987 \pm 0.11745$
	& $3.373 \pm 0.017$ & $3.78 \times 10^{-2}$ &   5.90 \\
4.5 & $1.07598 \pm 0.00699$ & 	
	$1.10246 \pm 0.07963$ & $1.09889 \pm 0.05829$
	& $3.527 \pm 0.017$ & $2.55 \times 10^{-2}$ & $-$     \\
5.5 & $1.08418 \pm 0.00752$ & 	
	$1.10409 \pm 0.08023$	& $1.09892 \pm 0.05758$
	& $3.681 \pm 0.019$ & $1.50 \times 10^{-2}$ & $-$     \\ \hline
\end{tabular}
\end{center}
\caption{
The optimal values of the three model parameters $s$, $a$ and $x$ obtained 
in three-parameter fits to $\nubar$ alone, 
as well as the corresponding mean neutron multiplicities $\nubar$,
together with the extracted uncertainties.  
The resulting values of $\chi^2_\nubar$ and $\chi^2_{\rm spectra}$
per degree of freedom are also given.}\label{a3paramfits}
\end{table}
\end{center}
\end{widetext}

\subsection{Determination of the model parameters}

Table \ref{a3paramfits} shows the optimal values 
and the associated uncertainties
for the three model parameters used in our fission calculations. 
These values have been obtained by fitting only to the evaluated $\nubar$ 
while ignoring the spectral data.
We have checked that fixing either $x$ or $a$, or both, in these fits
lead to equivalent results for all values of $E_n$.

The correlation coefficients between these model parameters are shown in 
Table~\ref{corrtable}.  If the parameters are uncorrelated, $C_{k k'} = 0$.
Correlated parameters lead to nonzero correlation coefficients.  If $C_{k k'}
>0$, $\alpha_k$ increases as $\alpha_{k'}$ increases.  On the other hand,
if $C_{k k'} < 0$, $\alpha_k$ increases as $\alpha_{k'}$ decreases.
The correlation coefficients between $s$ and $a$, $C_{s \, a}$, are relatively 
large and positive while those between $s$ and $x$, $C_{s\, x}$, are large 
and negative, suggesting strong correlations between these pairs of parameters.
The correlation coefficients between $a$ and $x$, $C_{a \, x}$, are close to
zero and fluctuate
in sign, signaling only a weak correlation between this pair of parameters.
In contrast, when the spectra are also included in the fits 
for $E_n \leq 3.5$ MeV, the correlation coefficients 
are all very close to $\pm 1$ in all cases, likely because the overlap in
parameter space that simultaneously reproduces $\nubar$ and the spectra 
is small.

\begin{table}[htpb]
\begin{center}
\begin{tabular}{cccc} \hline
$E_n$ (MeV) & $C_{s \, a}$ & $C_{s \, x}$ & $C_{a \, x}$ \\ \hline
0.5 & 0.608 & -0.569 & 0.0156 \\
1.5 & 0.611 & -0.561 & 0.0042 \\
2.5 & 0.465 & -0.776 & 0.0212 \\
3.5 & 0.464 & -0.766 & 0.0441 \\
4.5 & 0.757 & -0.569 &-0.0053 \\
5.5 & 0.693 & -0.480 &-0.0130 \\ \hline
\end{tabular}
\end{center}
\caption{The correlation coefficients (see Eq.~(\ref{sigmakk}))
for the three parameters $s$, $a$ and $x$ fitted to $\nubar$ alone.}
\label{corrtable}
\end{table}

The experimental values for the total average neutron multiplicity place 
remarkably stringent constraints on the value of the model parameter $s$
while more room is left for variations of $a$ and $x$.
Specifically, 
changing the tip separation distance scale factor $s$ by only $1\%$ 
(keeping $a$ and $x$ fixed) changes $\nubar$ by $1.8\%$, far outside the 
experimental uncertainty.  A change in $s$, see Eq.~(\ref{sparam}), results in 
a change in the average TKE, Eq.~(\ref{TKEdef}), of less than 0.5 MeV.  Thus
$\overline \nu$ is very sensitive to the balance between the kinetic and 
excitation energies.  On the other hand, $\overline \nu$ is less sensitive
to the partition of the excitation energy between the light and heavy fragments
since changing $x$ by 5\%\ (keeping $a$ and $s$ fixed)
changes $\nubar$ by only $0.5\%$.
Finally, $\overline \nu$ is least sensitive to changes in $a$ which modifies
the fragment temperature, predominantly affecting the low energy part of the
neutron spectrum.  Changing $a$ by 5\%\ (keeping $s$ and $x$ fixed) 
changes $\nubar$ by only $0.3\%$.

Table \ref{mdl2} shows results calculated by fitting to both 
$\overline{\nu}$ and the prompt neutron
spectra. (We do not show the 4.5 and 5.5 MeV results again since there
are no published spectra at these energies.)
When the spectral data are included in the fit the
agreement with these data and the evaluated $\overline{\nu}$ is poor. If we
had confidence in the spectral data, this would be a formal indication
that our model was incorrect or that uncertainties in $\overline{\nu}$ were
underestimated. Inconsistencies in the spectral data (see Sect.~\ref{spectra}) 
make either conclusion difficult. Some sets (particularly 
those of Alexandrova \cite{AleksandrovaPu}, which make the largest
contribution to the spectral $\chi^2$) are inconsistent with other sets, and, 
in a number of cases, uncertainties conservatively estimated.
In addition, the relative normalization, while determined from fitting to
a Watt spectrum and used only for scaling purposes, may increase the
relative $\chi^2$ for some data sets, possibly including the Aleksandrova
sets which are only available for $E > 1.5$ MeV.  Indeed, since these sets give
the largest contribution to the total $\chi^2$, eliminating them can change the
optimal parameter values, 
while removing one or more of the other sets has little to no effect.
For these reasons, we did not use the spectral data 
to obtain our final evaluation.
In addition, as discussed in more detail later, there are indications from
$^{235}$U measurements that more neutrons are emitted from the light fragment
than are from the heavy fragment ($x > 1$) \cite{NishioPu}.  
The fit at $E_n$=0.5~MeV shown in Table~\ref{mdl2} is consistent
with $x = 1$, giving $\overline \nu_L \approx \overline \nu_H$.

\begin{widetext}
\begin{center}
\begin{table}
\begin{center}
\begin{tabular}{ccccccc} \hline
$E_n$ (MeV) & $s^0$ & $a^0$ & $x^0$ & $\overline \nu$ 
& $\chi^2_{\overline \nu}$ & $\chi^2_{\rm spectra}/N$ \\ \hline
0.5 & ~$1.05705 \pm 0.00173$~ & ~$0.96754 \pm 0.02236$~ 
	& ~$1.00523 \pm 0.00574$~
& ~$2.961 \pm 0.007$~ & 0.76 & 13.72 \\
1.5 & $1.04573 \pm 0.00742$ & $0.97291 \pm 0.03424$ & $1.18356 \pm 0.05142$
& $3.078 \pm 0.020$ & 0.43 &  23.77 \\
2.5 & $1.05485 \pm 0.00602$ & $0.99909 \pm 0.04221$ & $1.18587 \pm 0.06274$
& $3.239 \pm 0.016$ & ~$0.0066$~ &  2.58 \\
3.5 & $1.05309 \pm 0.00657$ & $0.98038 \pm 0.03839$ & $1.21052 \pm 0.05293$
& $3.364 \pm 0.013$ & 0.24 &  4.61 \\ \hline
\end{tabular}
\end{center}
\caption{The optimal values $s^0$, $a^0$ and $x^0$
obtained in three-parameter fits to the spectra and $\overline \nu$.  The 
corresponding values of $\overline \nu$ are also shown.  
The resulting $\chi^2$ values for 
$\overline \nu$ and the spectra are given separately.}
\label{mdl2}
\end{table}
\end{center}
\end{widetext}

\subsection{The prompt neutron spectrum}

\begin{figure}[t]
\begin{center}
\includegraphics[width=3.1in]{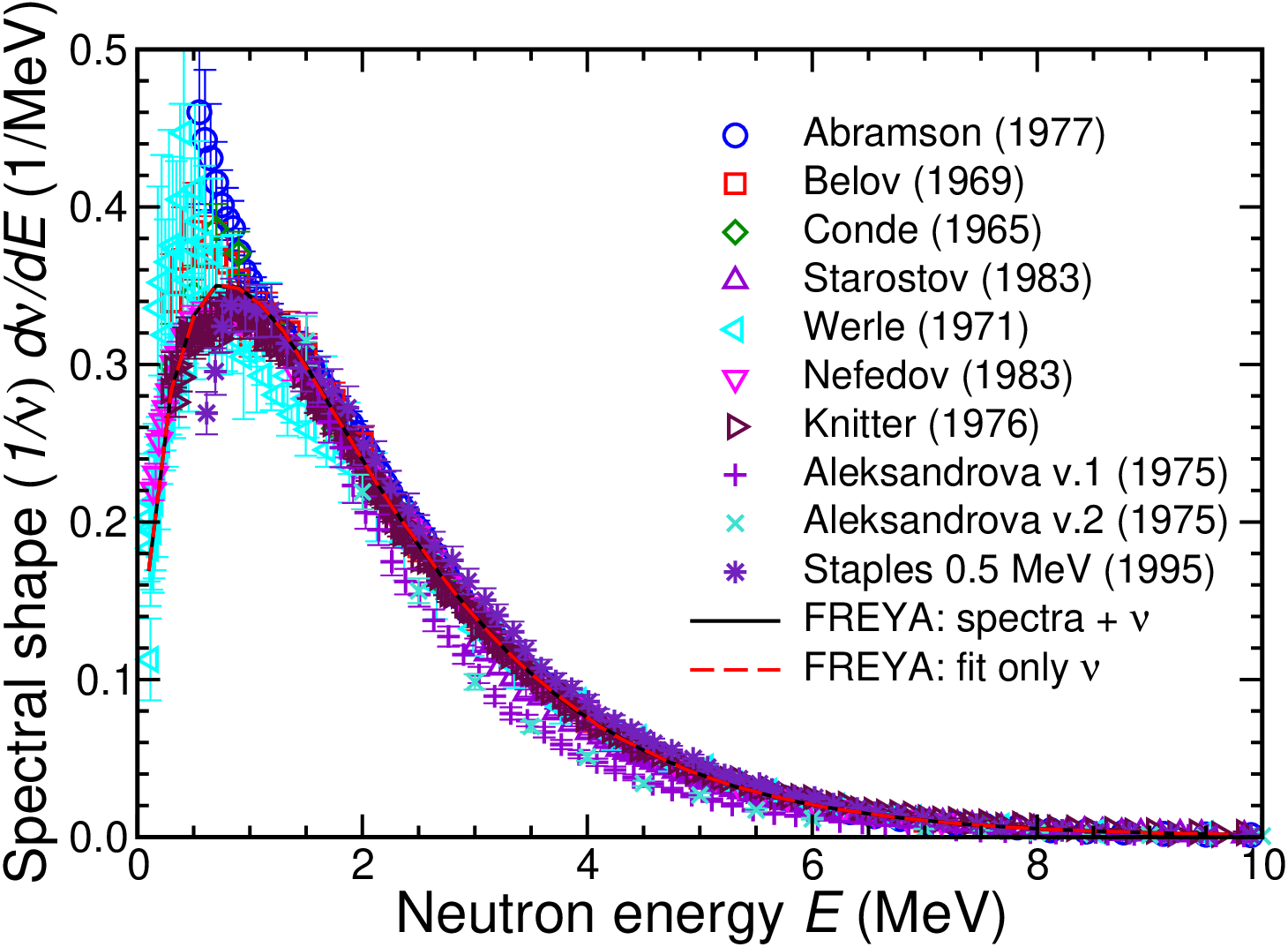}
\includegraphics[width=3.1in]{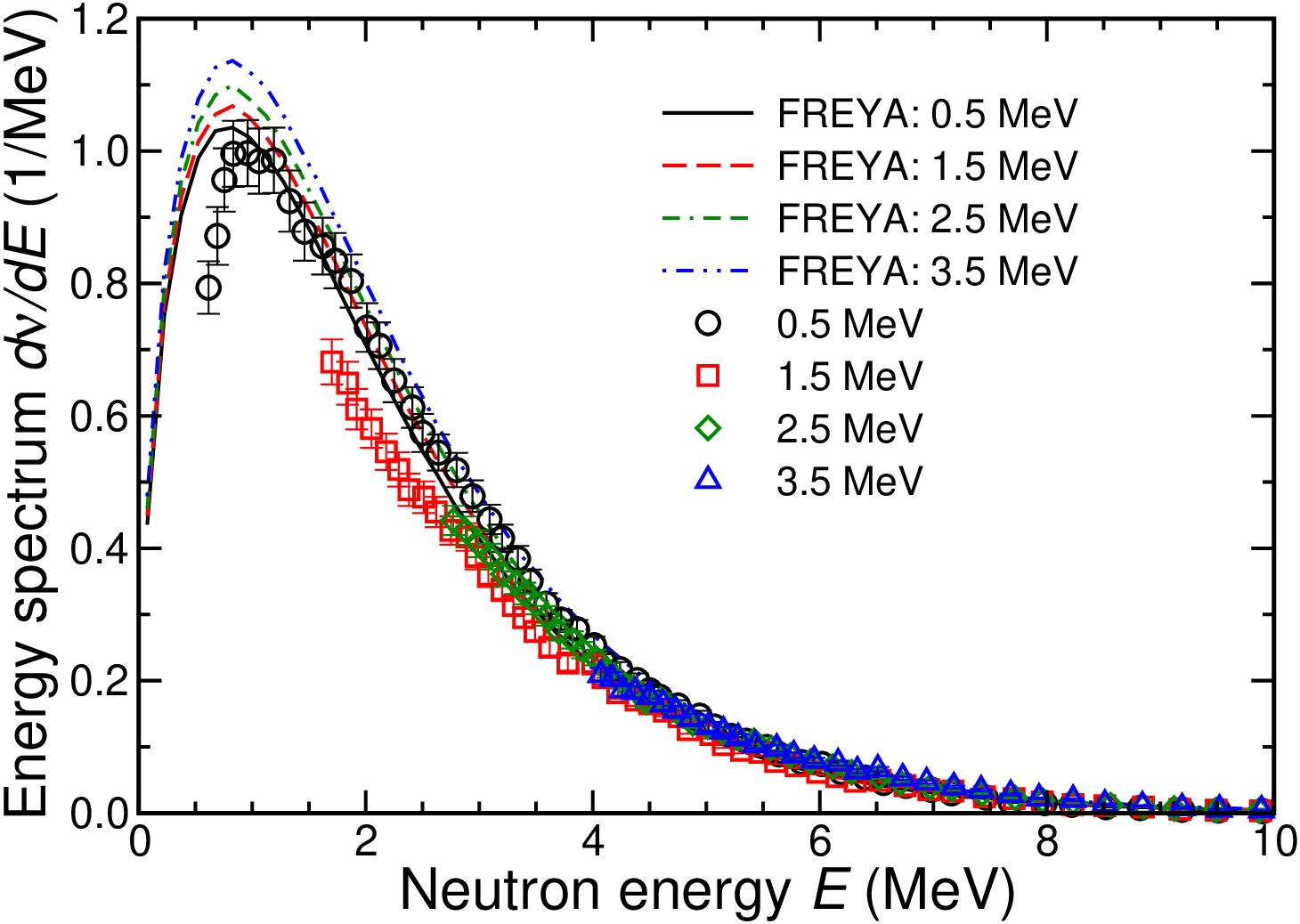}
\end{center}
\caption{(Color online)
The measured prompt neutron spectra are compared to our fit results.
The comparison to the low energy results from 
Refs.~\protect\cite{CondePu,BelovPu,WerlePu,KnitterPu,AleksandrovaPu,%
AbramsonPu,NefedovPu,StarostovPu,StaplesPu} ({\em upper panel}) are of the 
normalized spectral shapes while the results at higher incident neutron
energies from Ref.~\protect\cite{StaplesPu} ({\em lower panel})
are compared to the spectral distributions themselves.}
\label{f:nspec-fits}
\end{figure}

A comparison between experimental data and our calculations of the prompt 
neutron spectrum is shown in Fig.~\ref{f:nspec-fits}. The top panel of this 
figure gives the spectral shape and shows all experimental data from 
Refs.~\cite{{CondePu,BelovPu,WerlePu,KnitterPu,AleksandrovaPu,AbramsonPu,NefedovPu,StarostovPu,StaplesPu}}.  Since the shape varies slowly with incident 
neutron energy, the calculations using parameters fit to $\nubar$ alone and
to $\nubar$ and the spectral data are practically indistinguishable on a 
linear scale. The bottom panel of Fig.~\ref{f:nspec-fits} shows only the
more recent Staples data from 0.5 to 3.5 MeV \cite{StaplesPu}. In this panel, 
the different spectra can be distinguished because they have been normalized 
to $\overline{\nu}$, which varies modestly with incident
neutron energy. 

Because \code\ cannot produce sufficient statistics at
the fine energy scale needed by typical spectral evaluations, high statistics
\code\ runs have been made to emphasize the low and high energy tails of the 
spectra.  To remove statistical noise,  Watt distributions are fit to the low 
($E<2$~MeV) and high ($E>4$~MeV) energy parts of the spectrum for each 
incident neutron energy.  A fine grid
is obtained in the intermediate part of the spectrum by interpolation.

\begin{figure}[b]
\includegraphics[width=3.1in]{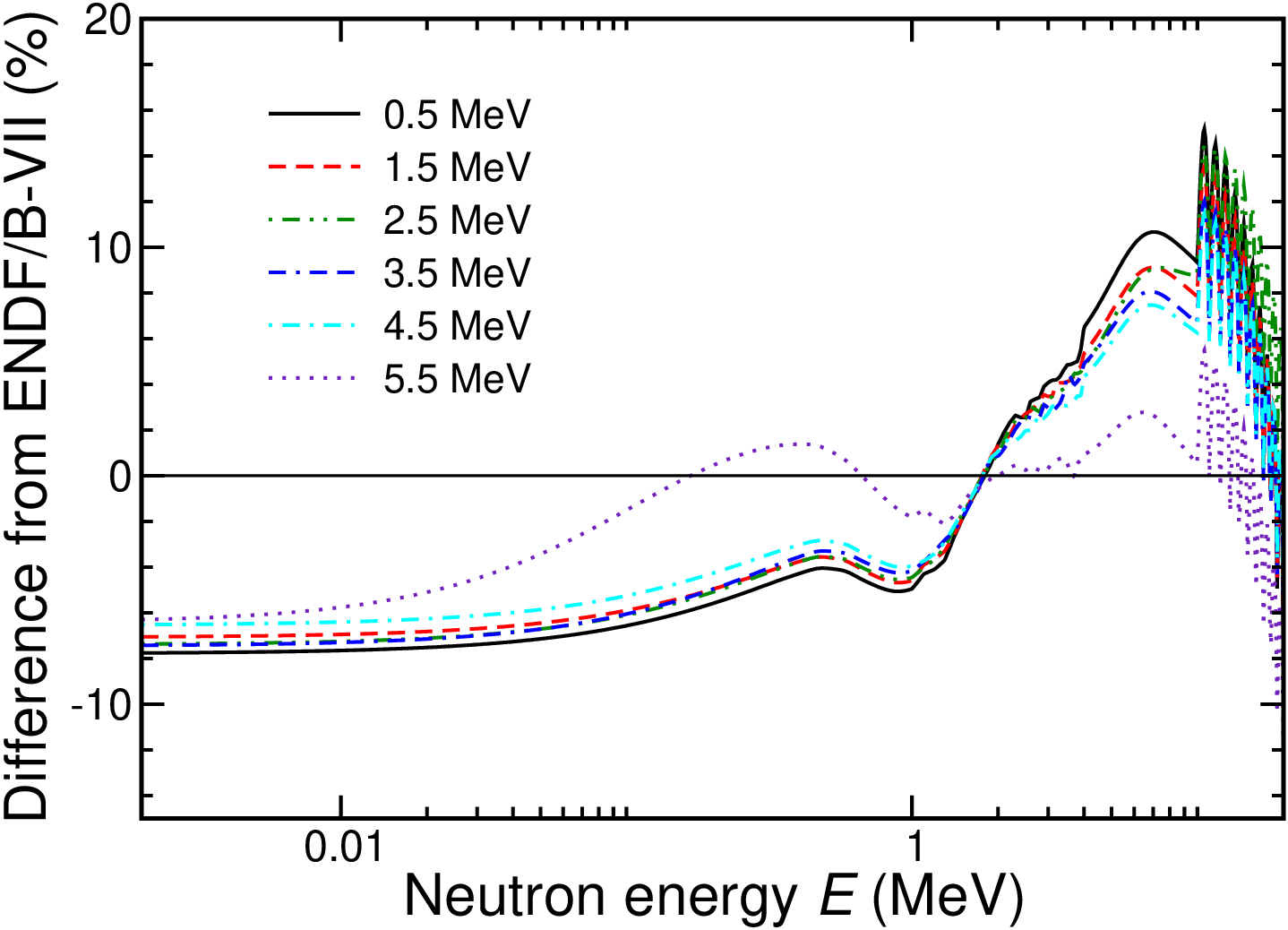}
  \caption{(Color online)
The percent difference between our evaluated spectra and that
of ENDF-B/VII for all six incident neutron energies considered.}
\label{endfdiff}
\end{figure}

Figure~\ref{endfdiff} gives the difference between the present calculations 
and the evaluations in ENDF/B-VII. 
Our spectra are systematically softer, giving lower mean neutron energies. 
This difference has important implications for criticality. 

\begin{figure}[t]
\includegraphics[width=3.1in]{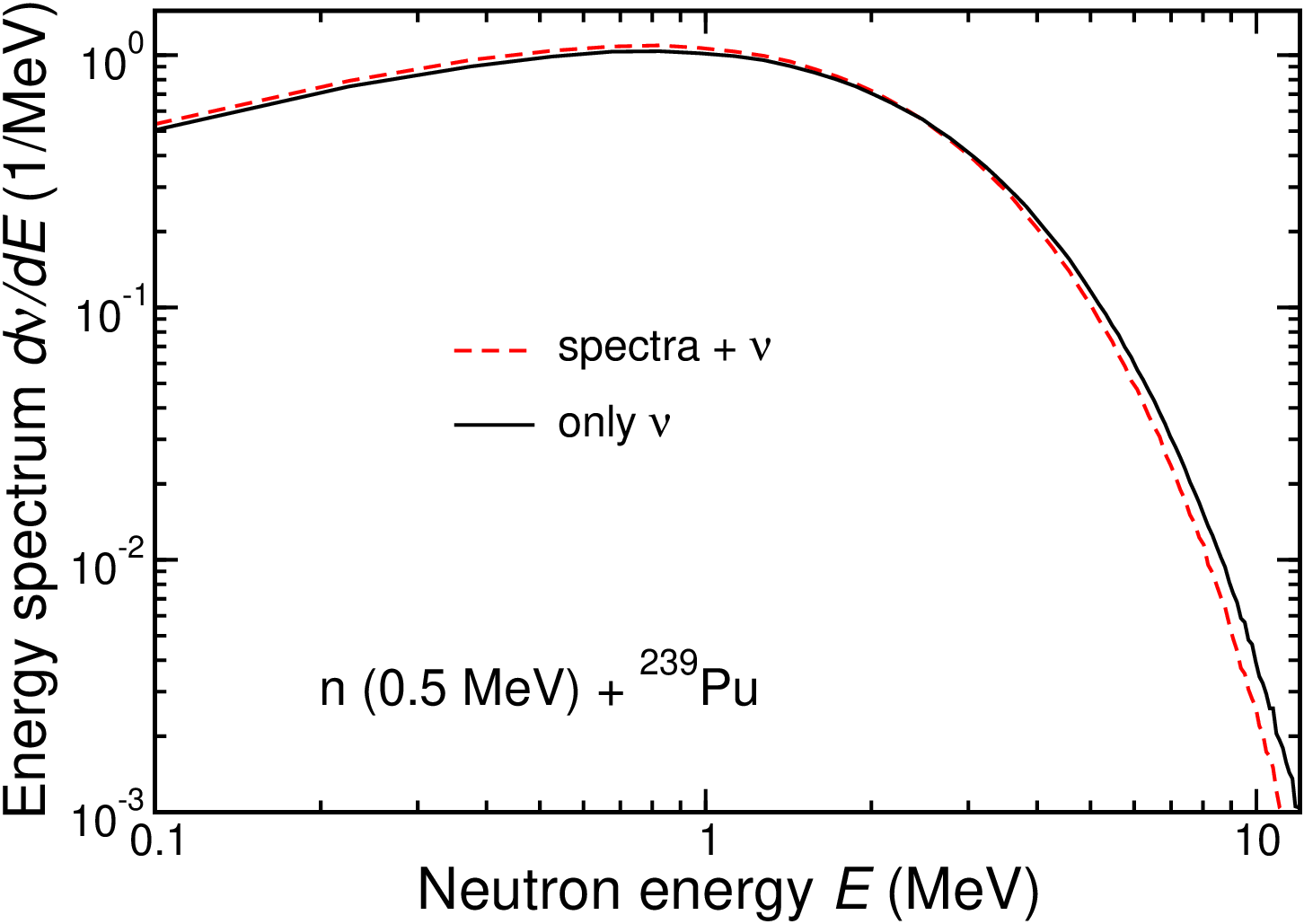}
  \caption{(Color online)
Prompt neutron spectra calculated in the laboratory frame as a function of 
outgoing neutron energy for 0.5 MeV incident neutron energies.  
The solid curve is
obtained by fitting $\overline \nu$ alone while the dashed curve
is fit to both the spectra and $\nubar$.}
\label{05MeVspectra}
\end{figure}

\begin{figure}[t]
\includegraphics[width=3.1in]{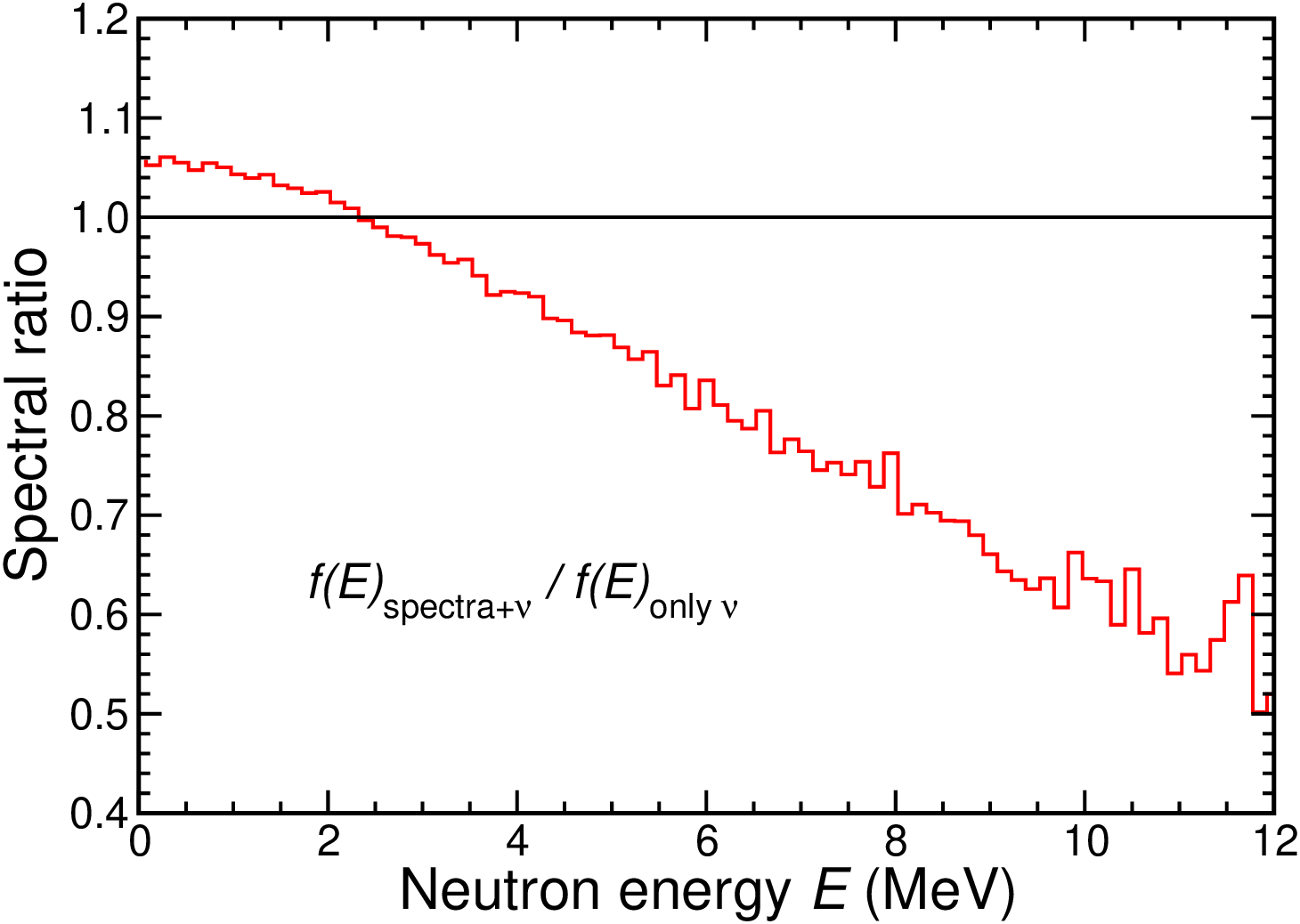}
  \caption{(Color online)
The ratio of the spectra obtained by fitting to $\nubar$ {\em and}
the spectral data relative to a fit based on $\nubar$ alone at $E_n=0.5$ MeV.}
\label{ratspnu}
\end{figure}

In the previous section, we argued that currrently available spectral data 
should not be used in the fission likelihood analysis. 
To illustrate the impact of these data on spectral calculations, we show
the difference between the fits without and with the spectral
data at $E_n = 0.5$~MeV, normalized to $\overline \nu$ on a
log-log scale, in Fig.~\ref{05MeVspectra}.  
The difference is largest in
the high-energy tail of the spectra where the fit to the spectra and
$\overline \nu$ is softer.  The ratio of the fits with and without the
spectral data are shown in Fig.~\ref{ratspnu}.  Below 2 MeV, the fit
with the spectral data is 1-2\% higher but by $E\approx 10$ MeV, it is
about 60\% lower than the spectral description with a fit to
$\overline \nu$ alone.  At higher energies the calculations grow
further apart but the ratios are statistics limited since, even with
1-2 million events \code\ does not fully populate the high energy tail
of the emission spectrum. 

\begin{figure}[t]
\includegraphics[width=3.1in]{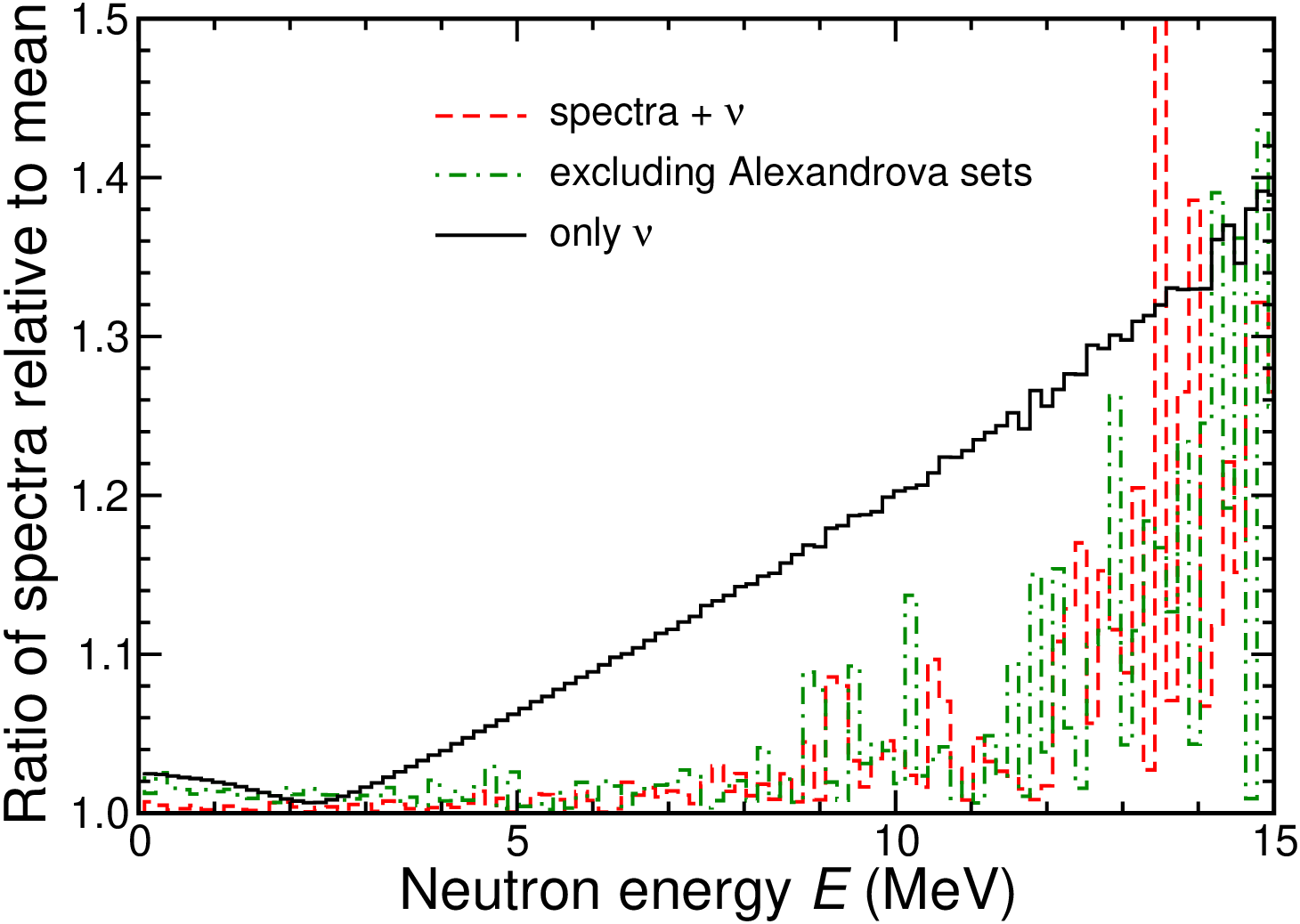}
  \caption{(Color online)
The spectral ratios (see text) for the three different 
analyses indicated.}
\label{relativeratios}
\end{figure}

We can compute the uncertainty in the spectra as well as in the 
employed values of the model parameters.  
Each particular set of model parameter values, $\{\alpha_k^{(m)}\}$, 
yields a different neutron spectrum $(d\nu/dE)^{(m)}$
so that the resulting ensemble of spectra can be subjected to a statistical 
analysis at each value of the neutron energy $E$,
yielding an average value of the neutron spectrum, $d\tilde{\nu}/dE$,
and an associated dispersion, $\sigma(d\nu/dE)$.
For $E_n=0.5$~MeV, Fig.~\ref{relativeratios}  shows the ratio between $d\tilde
{\nu}/dE+\sigma(d\nu/dE)$ and $d\tilde{\nu}/dE$.
This spectral ratio provides an indication of the relative 
uncertainty on the spectrum at each energy.
With the fits to $\overline \nu$ alone, the calculated uncertainty is less 
than 5\% for $E < 4$ MeV and less than 2\% for $E<2$ MeV, much smaller than
the spread in the data depicted in Fig.~\ref{f:nspec-fits}.  The uncertainty
increases approximately linearly with $E$ for
$E>2.5$ MeV, reaching $\approx 40$\% at 15 MeV.  We have also shown the
relative uncertainty with all spectra included in the fit as well as that 
obtained by leaving out the two spectra with the largest $\chi^2$.  Both of
these give small but noisy uncertainties, suggesting that result is not a
true measure of the calculated 
uncertainty in this case and that the spectral uncertainty as shown is rather
random.  The noisiness of the combined fits is due to the difficulty of
obtaining a combination of parameter values that simultaneously minimizes 
$\chi_{\overline \nu}^2$ and $\chi_{\rm spectra}^2$.

It is instructive to consider the correlations between the spectral strength
at different energies.
The evaluation of the corresponding covariance (see Eq.~(\ref{sigmaij}))
is complicated by the fact that the observables considered,
specified energies of emitted neutrons, form a continuum.
In practice, it is convenient to consider discrete energy bins 
(so the observable $\alpha_k$ represents the mean number of neutrons emitted 
with a kinetic energy in the bin $k$ centered at the energy value $E_k$.
Using Eq.~(\ref{sigmakk}), 
we may then calculate the corresponding covariance matrix
\beq
\tilde{\sigma}(E_k,E_{k'})\ =\ 
	\prec(E_k-\tilde{E}_k)(E_{k'}-\tilde{E}_{k'})\succ\ .
\eeq

\begin{figure}[t]
\includegraphics[width=3.1in,angle=-90]{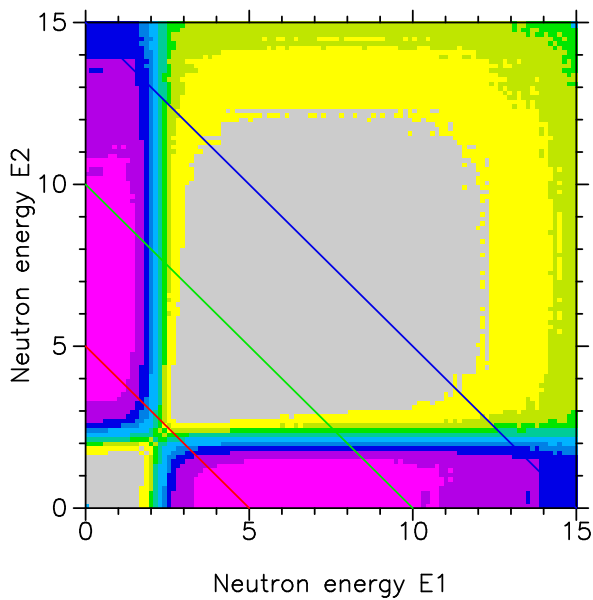}
  \caption{(Color online)
The correlation coefficients, $C(E_1,E_2)$, for the spectral
strength of the evaporated neutrons (see Eq.~(\protect\ref{C12})),
as obtained from the statistical analysis at $E_n=0.5$~MeV
when only $\nubar$ is considered in the fits.
Figure~\protect\ref{corE12} shows cuts along the three indicated lines 
of constant total energy.}
\label{corrmat}
\end{figure}

\begin{figure}[b]
\includegraphics[width=\columnwidth]{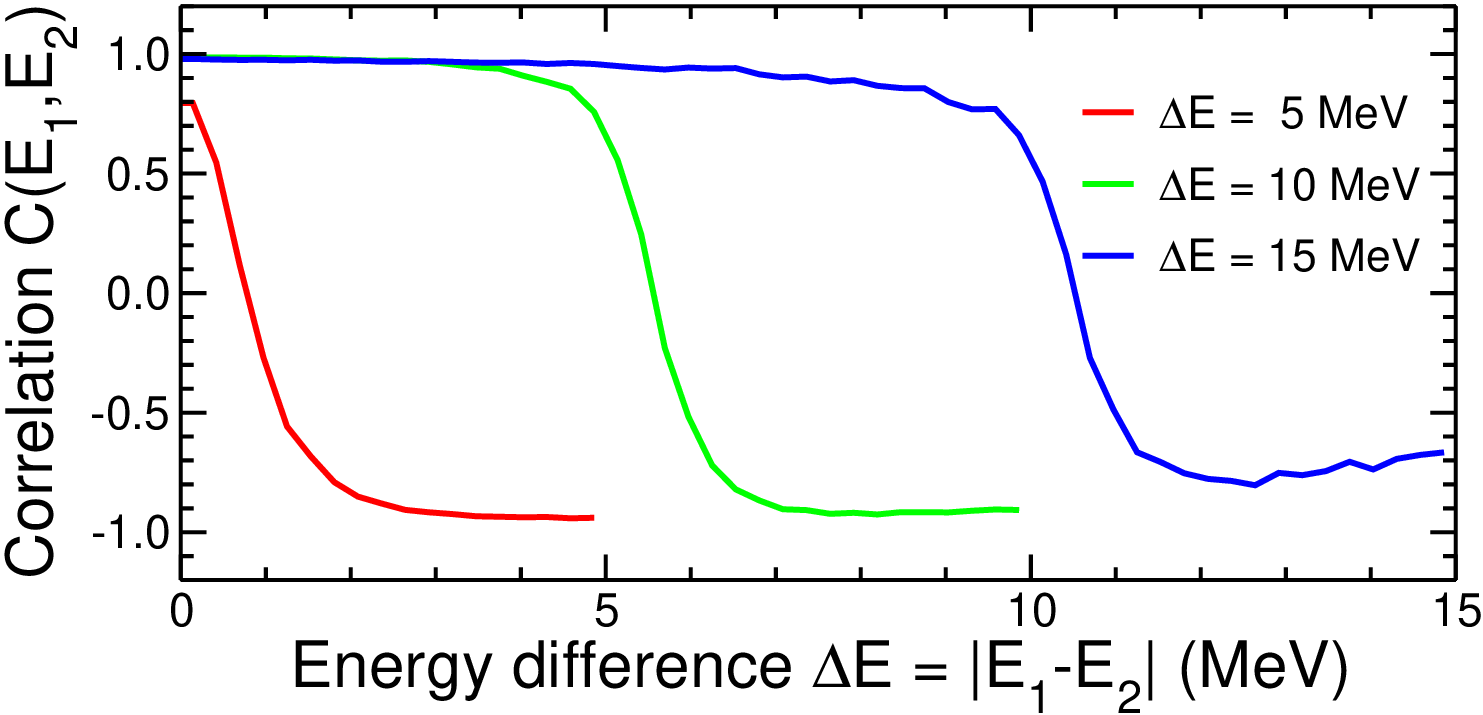}
  \caption{(Color online)
The spectral correlation coefficients, $C(E_1,E_2)$, along 
the three lines of constant combined energy indicated in Fig.~\ref{corrmat}.}
\label{corE12}
\end{figure}

However, it is important to recognize that for continuous observables,
the above matrix function is singular along the diagonal \cite{BurgioNPA529},
\beq
\tilde{\sigma}(E_k,E_{k'})\ =\ \tilde{\sigma}^2_{E_k}\,\delta(E_k-E_{k'})
	+\tilde{\sigma}_{E_kE_{k'}}\ ,
\eeq
where $\tilde{\sigma}^2_{E_k}$ is the variance in the differential yield
at the specified energy $E_k$, while $\tilde{\sigma}_{E_kE_{k'}}$ expresses
the correlation between the differential yields at two {\em different}
energies $E_k$ and $E_{k'}$.
To obtain this latter quantity, we must first remove the singular part.
This can be readily accomplished when the observable has been discretized
by simply replacing the diagonal elements in $\tilde{\sigma}(E_k,E_{k'})$
by values obtained by interpolating between the near-diagonal elements.
The resulting correlation coefficient,
\beq\label{C12}
C(E_k,E_{k'})\ =\ \tilde{\sigma}_{E_k E_{k'}}/
	[\tilde{\sigma}_{E_k}\tilde{\sigma}_{E_{k'}}]\ ,
\eeq
is then regular.
It is displayed in Fig.~\ref{corrmat} 
for the ensemble obtained for $E_n = 0.5$ MeV by fitting to $\nubar$ alone.
Figure~\ref{corE12} shows cuts at constant total neutron energy,
$E_k+E_{k'}$.
Similar results are found for all other incident energies considered.  

When the model parameters are varied,
the spectral shapes tend to pivot around $E\approx2$~MeV.
Consequently, when both neutron energies lie on the same side of this value,
the differential changes are in phase and the correlation coefficient
is close to one.0
The changes are in opposite directions 
when the two energy values are on opposite sides of the pivot energy.
%
By contrast, when the spectral data are included in the fits, 
the correlation coefficients vary widely
between $+1$ and $-1$ in no apparent pattern.  

When the number of \code\ events included in the $\nubar$-only fits 
at $E_n=0.5$~MeV 
is increased by a factor of five,
the fitted model parameter values change by less than one standard deviation.
When the spectra are also included in the fits, 
the resulting change in the fitted parameter values
increases $\chi^2_{\nubar}$ from 0.75 to $\approx$15 without significantly 
improving the spectral fits.  
Moreover, while the fluctuations in the energy correlation coefficients
decrease somewhat when the larger event samples are used,
they do not disappear.

As is the case for the model parameters, there are uncertainties in
the spectral calculations. If the model is
qualitatively wrong, and the right spectral form cannot be obtained by simply
changing the parameter values, then the spectral uncertainties are not
correct. 
To explore this we performed several additional variations on the model. 
In Sec.~\ref{model} we saw that a model that employs a constant tip separation
$d$, independent of the specific binary partition, 
reproduces neither the total kinetic energy data nor the neutron yield as a
function of fragment mass. 
Nevertheless, a constant $d_{LH}$ yields better agreement 
with the $\nubar$-only fit in Fig.~\ref{05MeVspectra} 
than with the fit that also includes the spectra.
Similarly, making the level 
density parameter independent of energy, $\tilde a = A/e_0$, 
changes the spectrum by less than 5\%\ at lower energies ($<5\,{\rm MeV}$)
and by less than 20\% at higher energies.
Fundamental microscopic calculations of 
fission could provide insight into the sensitivity of the spectrum to changes
in the parameters, leading to better estimates of the spectral uncertainties. 

Critical assemblies, which are designed to determine the conditions
under which a fission chain reaction is stationary,
provide an important quality check on the spectral evaluations.  
The key measure of a critical assembly is 
the neutron multiplication factor $k_{\rm eff}$ 
(often denoted as the $k$ eigenvalue).
When this quantity is unity, the assembly is exactly critical,
{\em i.e.}\ the net number of neutrons resulting from each neutron-induced
fission event is one on the average.
(This number is the difference between the number of neutrons emitted 
during the fission process and those lost to absorption and escape.)
The degree of criticality of a particular assembly depends on 
the multiplicity of prompt neutrons, their spectral shape,
and the $(n,f)$ induced-fission cross section.

Plutonium criticality is especially sensitive to the prompt
neutron spectrum because the $^{239}{\rm Pu}(n,f)$ cross section 
rises sharply between $E_{n} = 1.5$ and 2 MeV. 
As a result, increasing the relative number of low-energy neutrons 
tends to decrease criticality, lowering $k_{\rm eff}$,
while increasing the number of neutrons having higher energy 
increases criticality. 

\begin{figure}[t]
\includegraphics[width=3.1in]{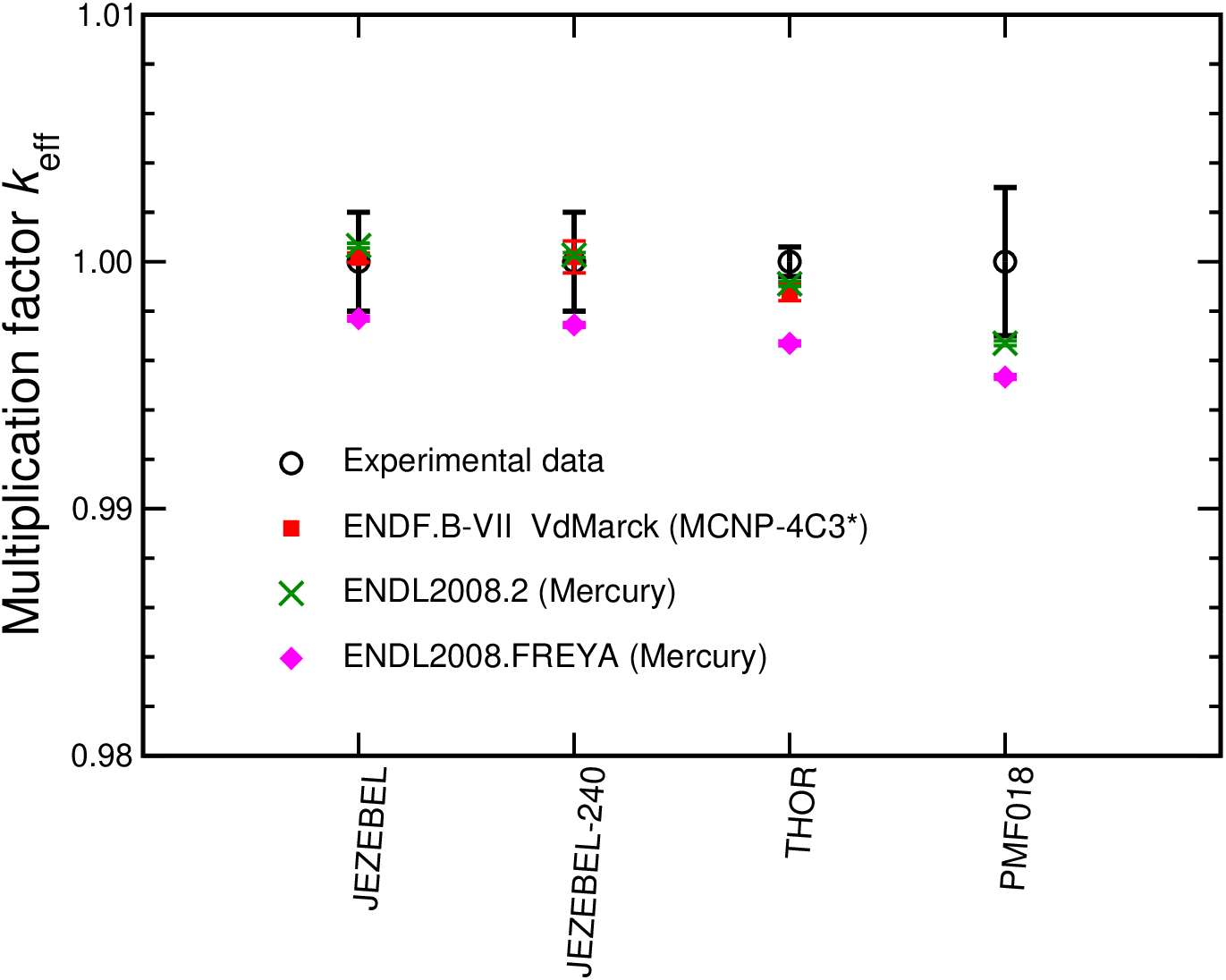}
  \caption{(Color online)
Calculated $k_{\rm eff}$ for several $^{239}$Pu
critical assemblies obtained using our fits for $0.5 \leq E_n \leq 5.5$ MeV
in the Mercury Monte Carlo.  The results are compared to those with
the standard ENDL2008.2 and ENDF-B/VII databases.}
\label{assembliesfig}
\end{figure}

Figure~\ref{assembliesfig} 
shows calculations of $k_{\rm eff}$ for different plutonium assemblies. 
Apart from the spectra, all data used in these calculations were taken 
from ENDF/B-VII. 
Overall there is good agreement with the measured values of $k_{\rm eff}$, 
though this new softer spectrum decreases the calculated values by about 0.003.
Since this is approximately 1.5 standard deviations away from the measurement,
there may be an indication that the Pu fission cross section
or neutron multiplicity is low by about a tenth of a percent. 
There appears to be room for some adjustment of the experimental
data since the uncertainties in the cross sections are about 1\%, 
while those in $\nubar$ are about 0.5\%.

\subsection{More exclusive observables}

\begin{figure}[t]
\begin{center}
\includegraphics[width=3.1in]{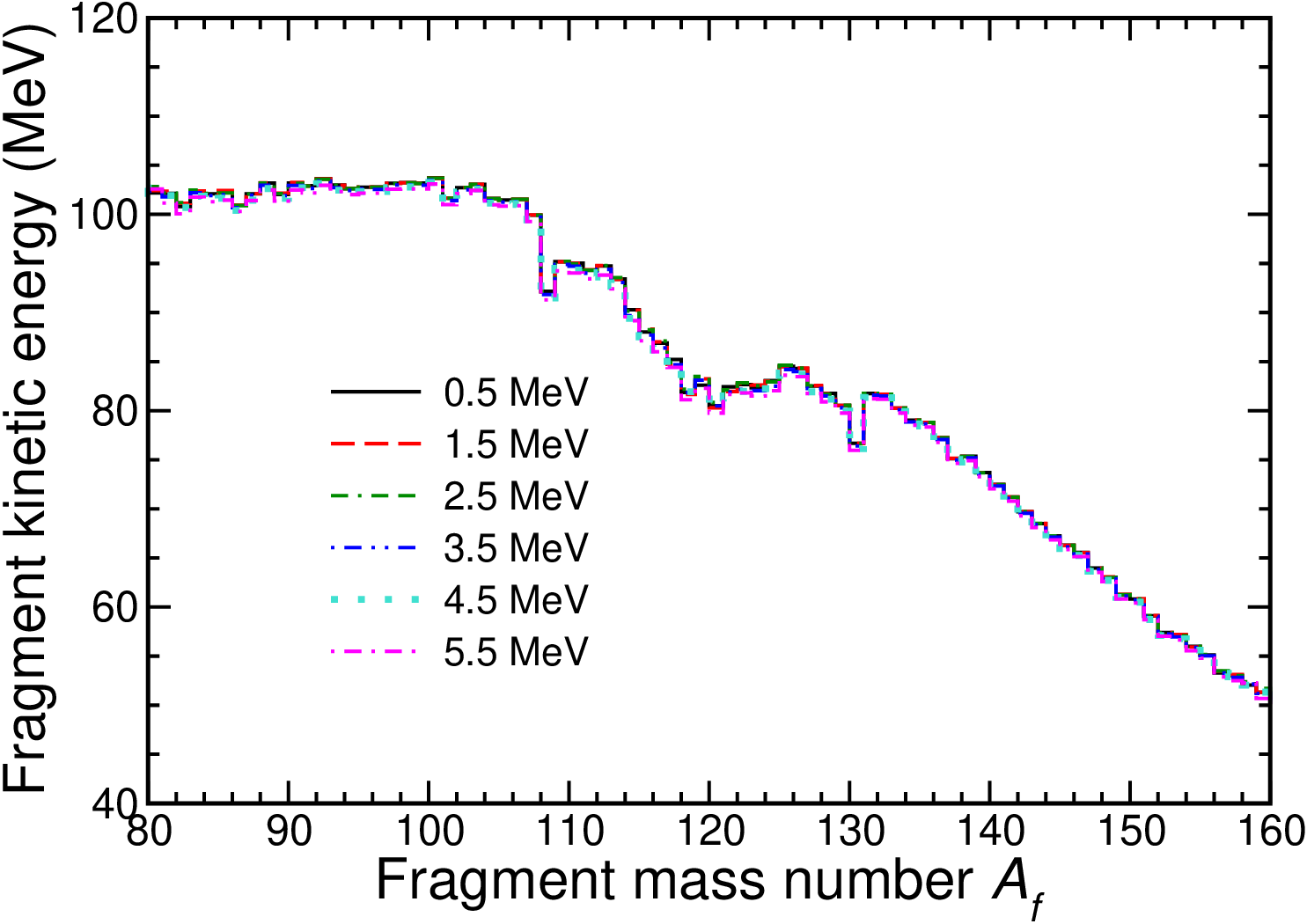}
\includegraphics[width=3.1in]{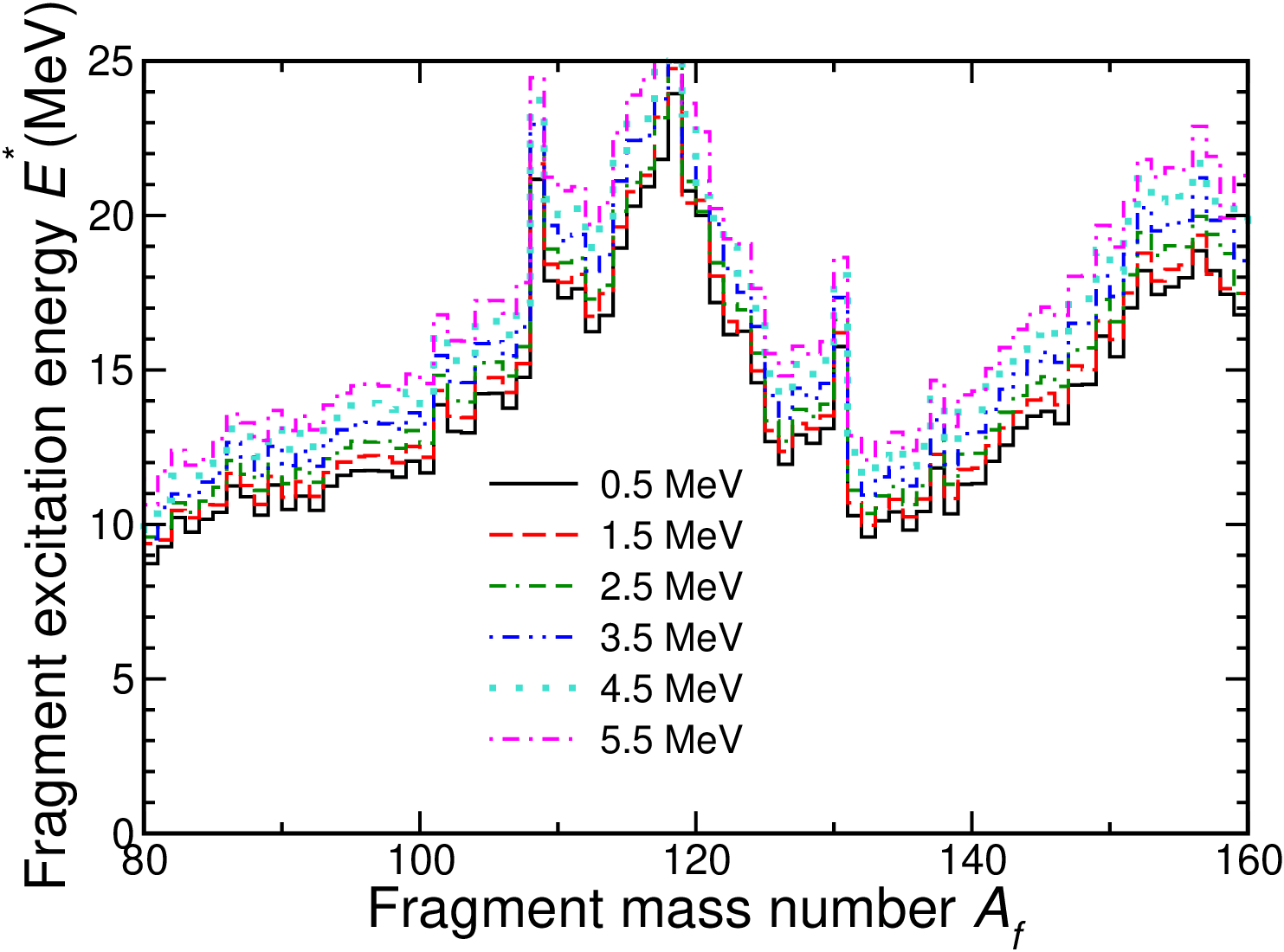}
\end{center}
\caption{(Color online)
The average kinetic energy of the fission fragments ({\em upper panel})
and their average excitation ({\em lower panel})
as a function of fragment mass number $A_f$ 
for $0.5 \leq E_n \leq 5.5$ MeV.}
\label{cap:tke-tee-edep}
\end{figure}

Though less important for understanding energy production,
more exclusive observables play a central role in the development of 
a comprehensive description of fission. Figure~\ref{cap:tke-tee-edep} 
shows calculations of fragment kinetic and excitation energies. 
Note that the fragment kinetic energies are almost independent of
the incident neutron energy.  Indeed, the  kinetic energy appears to decrease
slightly with energy, as may be expected since $s$ increases.  This may at
first appear surprising but the Coulomb approximation to the total kinetic
energy in Eq.~(\ref{TKEdef}) is independent of the incident neutron energy.
These results are consistent with measurements made with $^{235}$U and 
$^{238}$U targets over a similar incident neutron energy range, $0.5 \leq E_n
\leq 6$ MeV \cite{Hambschpriv} and $1.2 \leq E_n \leq 5.8$ MeV \cite{Vives238}
respectively.  In both cases, the average TKE, $\overline{\rm TKE}$, changes
less than 1 MeV over the entire energy range.  Ref.~\cite{Vives238} also
shows that, while the mass-averaged TKE is consistent with near energy 
independence, higher energy incident neutrons typically give more TKE to
masses close to symmetric fission and somewhat less TKE for $A_H > 140$.
The slight increase in TKE close to symmetric fission of $^{238}$U is not
unexpected since the symmetric contribution to $Y(A)$ increases with incident 
neutron energy.  Since such detailed TKE information is not available for
neutrons on $^{239}$Pu, we have therefore chosen to use a constant scale factor
at each energy.

\begin{figure}[t]
\begin{center}
\includegraphics[width=3.1in]{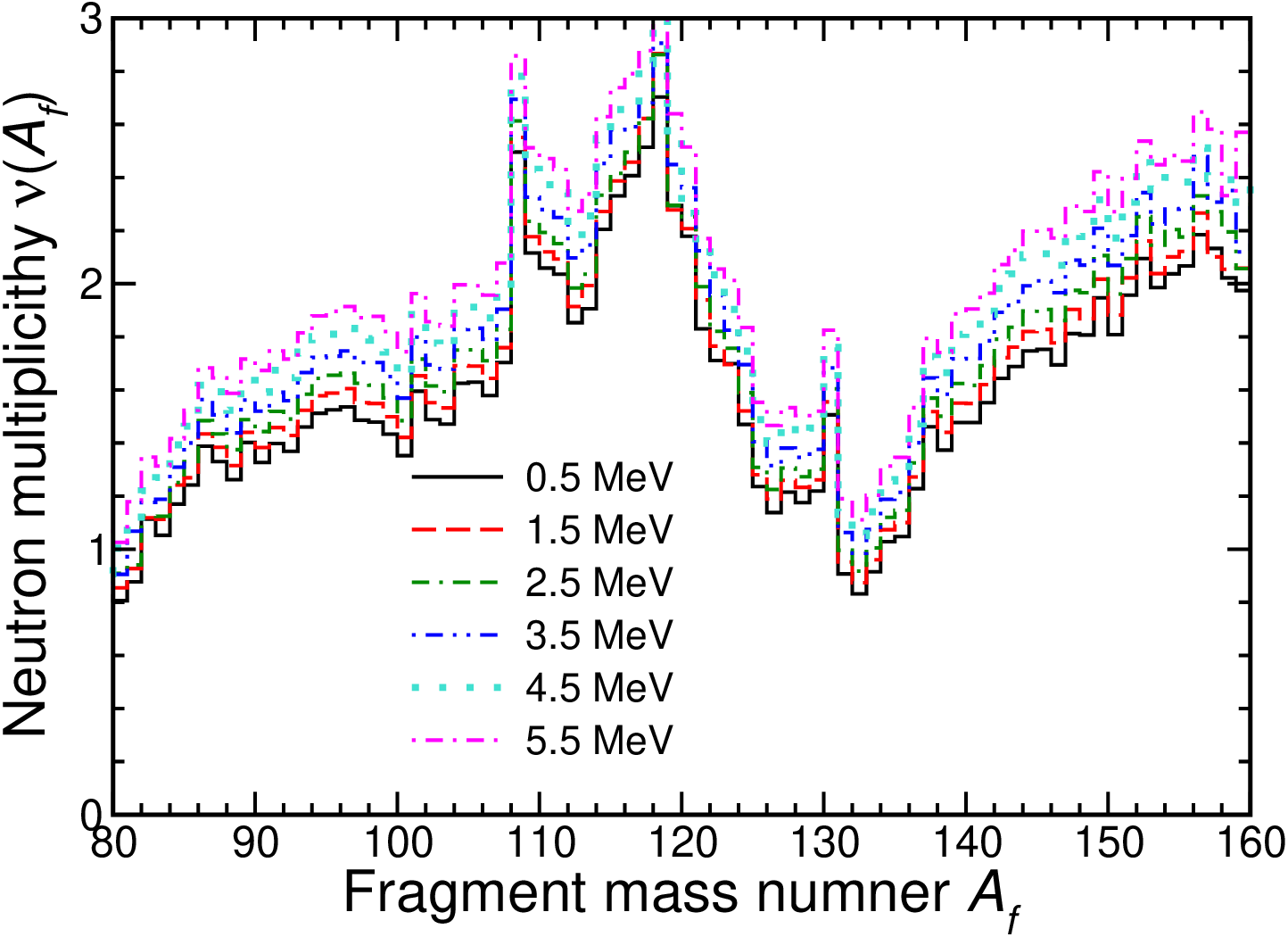}
\includegraphics[width=3.1in]{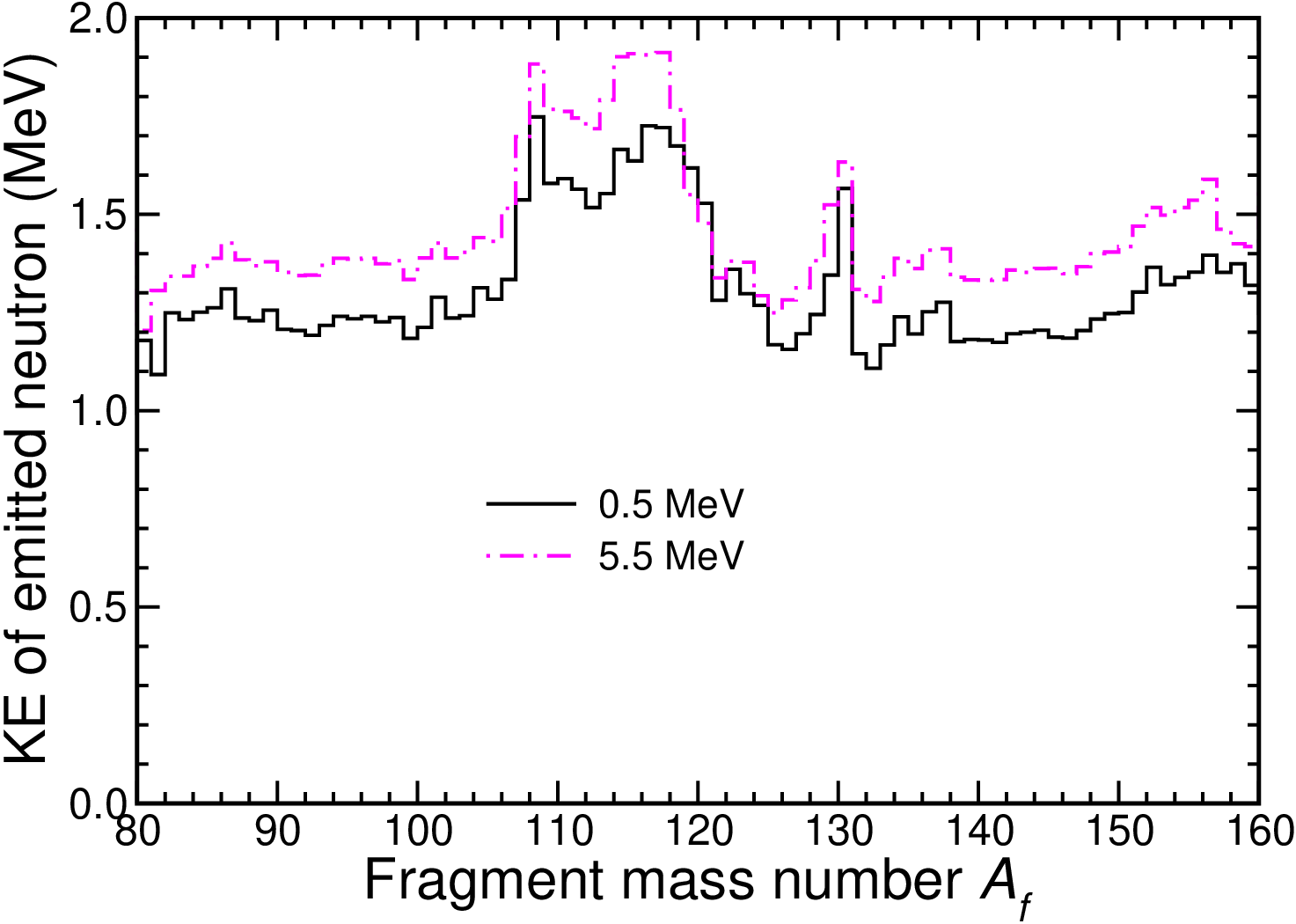}
\end{center}
\caption{(Color online)
The average neutron multiplicity 
for $0.5 \leq E_n \leq 5.5$ MeV ({\em upper panel}) and
the mean kinetic energy of the evaporated neutrons ({\rm lower panel})
for $E_n = 0.5$ and 5.5 MeV,
as functions of the fission fragment mass number $A_f$.}
\label{cap:nuvsa-keneut-edep}
\end{figure}

The constancy of the fragment kinetic energy as a function of $E_n$ allows 
the energy of the incident neutron to be converted into excitation energy.
The increase of $E^*$ with $E_n$ is fairly monotonic over all $A_f$, 
see the right-hand side of Fig.~\ref{cap:tke-tee-edep}.  
It appears, however, that the slope of $\nu(A_f)$ for $A_f > 120$ increases 
somewhat faster with $E_n$ than for $A_f < 120$, as shown on the 
left-hand side of Fig.~\ref{cap:nuvsa-keneut-edep}.  This scenario 
is consistent with washing out the sawtooth pattern
of $\nu(A_f)$ with increasing neutron energy \cite{Vandenbosch}.  See 
Table~\ref{avenubars} for the average neutron multiplicity for the light
and heavy fragments as well as the sum.
The associated multiplicity dispersions,
$\sigma_\nu=[\langle \nu \rangle^2 - \nubar^2]^{1/2}$, are also given.
Since $\nubar$ is used to determine the values model parameters, 
it may be preferable to use a different (and more exclusive) observable 
to check whether a given model parameter set is preferred over another. 
 A better choice is the average neutron multiplicity
and average neutron energies from the individual fragments.  
There are some limited data on thermal neutron-induced fission of $^{235}$U 
\cite{NishioU} and spontaneous fission of $^{252}$Cf \cite{VorobyevCf}
which suggest that the light fragment
emits more neutrons than the heavy fragment, 40\% more for $^{235}$U 
\cite{NishioU} and 20\% more for $^{252}$Cf \cite{VorobyevCf}.  
Our results for 0.5 MeV, shown
in Table~\ref{avenubars}, give a relative difference in $\overline \nu$
between the light and heavy fragments of about 20\% for $x \sim 1.1$.
Fits to $\overline \nu$ and the spectral data rather than $\overline \nu$
alone give $\overline \nu_L \approx
\overline \nu_H$ for $E_n \leq 0.5$ MeV, seemingly excluded by these
measurements, if the same is true for Pu.

A more sensitive neutron observable is the kinetic energy
of an evaporated neutron.
The lower panel in Fig.~\ref{cap:nuvsa-keneut-edep} shows the 
average kinetic energies of the
emitted neutrons as a function of fragment mass for the lowest and highest
incident energies studied (0.5 and 5.5 MeV).   The average kinetic energy of
the emitted neutrons is almost constant with $A$ except in the region
$110 < A < 140$ where it increases.  The dip in TKE occurs in the symmetric
region, making more energy available for neutron emission, 
resulting in more and faster prompt neutrons.

\begin{table}[hbp]
\begin{center}
\begin{tabular}{ccccccc} \hline
$E_n$ (MeV) & $\nubar$ & $\sigma_{\overline \nu}$ & $\overline \nu_L$ 
& $\sigma_{\overline \nu_L}$ & $\overline \nu_H$ & $\sigma_{\overline \nu_H}$ 
\\ \hline
0.5 & ~2.947~ & ~1.381~ & ~1.604~ & ~0.723~ & ~1.343~ & ~0.676~ \\
1.5 & 3.090 & 1.400 & 1.685 & 0.755 & 1.405 & 0.704 \\
2.5 & 3.244 & 1.424 & 1.761 & 0.783 & 1.483 & 0.738 \\
3.5 & 3.376 & 1.443 & 1.828 & 0.806 & 1.548 & 0.767 \\
4.5 & 3.530 & 1.466 & 1.905 & 0.833 & 1.624 & 0.801 \\
5.5 & 3.683 & 1.499 & 1.983 & 0.863 & 1.699 & 0.836 \\
\hline
\end{tabular}
\end{center}
\caption{The mean combined neutron multiplicities $\nubar$
as well as the mean multiplicities of neutrons emitted from
either the light or the heavy fragment, $\nubar_L$ and $\nubar_H$,
together with the associated dispersions.}\label{avenubars}
\end{table}

\begin{figure}[t]
\includegraphics[width=3.1in]{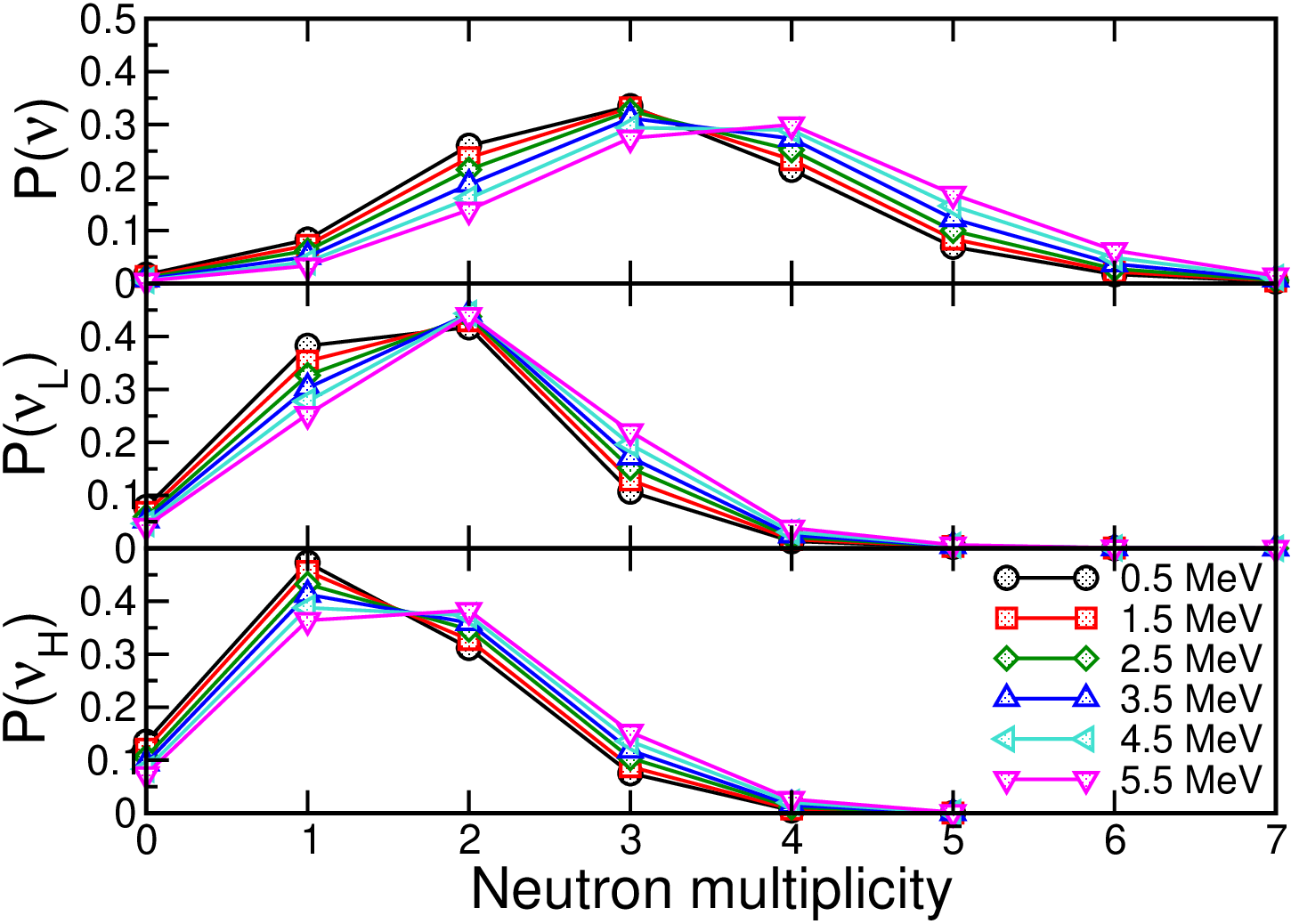}
  \caption{(Color online)
The normalized neutron multiplicity distribution obtained with
\code\ for both fragments (top), the light fragment (middle) and the heavy
fragment (bottom).  Results are shown for $0.5 \leq E_n \leq 5.5$ MeV.}
\label{pofnu}
\end{figure}

In Fig.~\ref{pofnu} we show the probability for a given neutron multiplicity,
$P(\nu)$, as a function of neutron number for all $E_n$.  
Along with the probability distribution
for emission from both fragments, we also show the distributions for the light
and heavy fragments separately.  

\begin{table}[t]
\begin{center}
\begin{tabular}{ccccccc} \hline
$E_n$   
& $\langle E^{L+H} \rangle$ & $\sigma_E^{L+H}$
& $\langle E^L \rangle$ & $\sigma_E^L$ 
& $\langle E^H \rangle$ & $\sigma_E^H$ 
\\ \hline
0.5 & ~2.054~ & ~1.625~ & ~2.313~ & ~1.755~ & ~1.750~ & ~1.418~ \\
1.5 & 2.088 & 1.654 & 2.346 & 1.787 & 1.785 & 1.448 \\
2.5 & 2.113 & 1.674 & 2.369 & 1.809 & 1.816 & 1.474 \\
3.5 & 2.140 & 1.698 & 2.397 & 1.836 & 1.828 & 1.496 \\
4.5 & 2.168 & 1.721 & 2.425 & 1.860 & 1.873 & 1.521 \\
5.5 & 2.198 & 1.746 & 2.455 & 1.883 & 1.905 & 1.546 \\
\hline
\end{tabular}
\end{center}
\caption{The average energy of neutrons emitted by either fragment and
by the light and heavy fragments separately, along with the associated 
dispersions, for various incoming neutron energies (all in MeV).
}\label{averageenergies}
\end{table}

Table~\ref{averageenergies} gives the average energies of the neutrons
emitted from the light fragment, the heavy fragment, or from either,
together with the associated variances,
for the incident neutron energies $E_n$.
The average energies increase with $E_n$ in all cases
and those coming from the light fragment tend to be more energetic
than those coming from the heavy one,
so we have  $\langle E^L \rangle >\langle E^{L+H}\rangle >\langle E^H\rangle$.
The variances exhibit the same hierarchy as the average energies
but increase more slowly with incident energy.  
The overall average energy $\langle E\rangle$ is similar to that obtained
for thermal neutron-induced fission of $^{235}$U and $^{252}$Cf(sf) 
found in Ref.~\cite{LemairePRC72}.

\section{Conclusion}

Our studies employ the recently developed a Monte-Carlo model,
\code, that simulates fission and the subsequent neutron and photon emission 
from the fragments on an event-by-event basis, 
maintaining energy and momentum conservation at each step 
in the production  and de-excitation of the fragments.
We have introduced three adjustable parameters, $s$, $a$, and $x$,
which modulate the separation between the tips of the fragments,
scale the level-density parameter for the fragments, and
modify the partition of energy between them, respectively.
These three model parameters were varied over an appropriate range and, for 
each particular set of values,
\code\ was used to generate a large sample of fission events
from which the resulting properties of the neutron spectra were extracted.
Each set of parameter values was assigned a likelihood weight 
based on the $\chi^2$ obtained from comparison with the measured mean 
multiplicity $\overline{\nu}$ and/or 
the measured differential neutron spectrum $d\overline{\nu}/dE$.
Mean values and covariances for both input parameters and quantities predicted
by the model were obtained through standard statistical techniques. 
This combination of the Monte-Carlo fission model with the likelihood weighting
presents a powerful tool for the evaluation of fission-neutron data. 

This procedure was applied to the analysis of neutron-emission data
for neutron-induced fission on $^{239}\textrm{Pu}$, from thermal
to 5.5 MeV incident energies. 
Although the approach taken and the nucleus studied
in this work are different, the results largely corroborate the findings
of Lemaire {\it et al.} \cite{LemairePRC72} in emphasizing the importance
of the initial conditions ({\it e.g.} the kinetic and excitation energies
of the fragments). Furthermore, our work underscores the effectiveness
of the measured $\overline {\nu}$ in constraining the model parameters,
more strongly even than the differential neutron-spectrum data. In
particular, it was found the the parameter controlling the tip separation
between fragments was by far the most important in reproducing the
experimental $\overline {\nu}$ values. In the end, fits of our model to
the $\overline{\nu}$ data alone ({\it i.e.}, excluding the 
differential-spectral data) were found to be more robust and were used to 
obtain the best model parameters.

We plan to apply this method to the prediction of neutron
emission properties in other actinides. 
However, in those cases where critical experimental data 
(such as kinetic energies and neutron multiplicities and spectra) 
are not available to constrain the \code\ calculations,
it may be necessary to invoke supplementary information from 
various theoretical models, such as Hartree-Fock-Bogoliubov 
or macroscopic-microscopic treatments.

\section*{Acknowledgements}

We wish to acknowledge many helpful discussions with 
D.A.~Brown, M.-A.~Descalle, D.~Gogny, E.~Ormand, P.~M{\"o}ller,
E.B.~Norman, W.J.\ Swiatecki, and P.~Talou.
This work was performed under the auspices of the 
U.S. Department of Energy by Lawrence Livermore National Laboratory under 
Contract DE-AC52-07NA27344 (RV, JP, WY), by Lawrence
Berkeley National Laboratory under Contract DE-AC02-05CH11231 (JR) 
and was also supported in part by the National 
Science Foundation Grant NSF PHY-0555660 (RV).

\begin{widetext}

\end{widetext}

\end{document}